\documentclass[fleqn,10pt]{wlscirep}
\usepackage[utf8]{inputenc}
\usepackage[T1]{fontenc}
\usepackage{lineno}
\usepackage{amsmath}
\usepackage{gensymb}
\usepackage{overpic}
\usepackage{xcolor}
\usepackage{soul} 

\title{Modelling Carbon Capture on Metal-Organic Frameworks with Quantum Computing}
\author[1]{Gabriel Greene-Diniz}
\author[1]{David Zsolt Manrique}
\author[2]{Wassil Sennane}
\author[3]{Yann Magnin \footnote[1]{Consultant, on the behalf of TotalEnergies S.E.}}
\author[2]{Elvira Shishenina \footnote[3]{current address \textit{BMW Group, New Technologies and China, 80788 Munich, Germany}}}
\author[2]{Philippe Cordier}
\author[3]{Philip Llewellyn}
\author[1]{Michal Krompiec}
\author[2]{Marko J. Ran\v{c}i\'{c}$^\dagger$}
\author[1]{David Mu{\~n}oz Ramo$^{\dagger \dagger}$}

\affil[1]{\textit{Cambridge Quantum Computing Ltd, 13-15 Hills Road, CB2 1NL, Cambridge, UK}}
\affil[2]{\textit{TotalEnergies, OneTech, One R\&D, 8 Boulevard Thomas Gobert, 91120 Palaiseau, France}}

\affil[3]{\textit{TotalEnergies, OneTech, CO$_2$ \& Sustainability R\&D, CSTJF - Avenue Larribau, 64018 Pau Cedex, France}}

\affil[$\dagger$]{e-mail: marko.rancic@totalenergies.com}
\affil[$\dagger \dagger$]{e-mail: david.ramo@cambridgequantum.com}

\begin{abstract}
Despite the recent progress in quantum computational algorithms for chemistry, there is a dearth of quantum computational simulations focused on material science applications, especially for the energy sector, where next generation sorbing materials are urgently needed to battle climate change. To drive their development, quantum computing is applied to the problem of CO$_2$ adsorption in Al-fumarate Metal-Organic Frameworks. Fragmentation strategies based on Density Matrix Embedding Theory are applied, using a variational quantum algorithm as a fragment solver, along with active space selection to minimise qubit number. By investigating different fragmentation strategies and solvers, we propose a methodology to apply quantum computing to Al-fumarate interacting with a CO$_2$ molecule, demonstrating the feasibility of treating a complex porous system as a concrete application of quantum computing. We also present emulated hardware calculations and report the impact of device noise on calculations of chemical dissociation, and how the choice of error mitigation scheme can impact this type of calculation in different ways. Our work paves the way for the use of quantum computing techniques in the quest of sorbents optimisation for more efficient carbon capture and conversion applications.
\end{abstract}

\begin{document}

\flushbottom
\maketitle

\thispagestyle{empty}
\section*{Introduction} \label{main_t} The capture of carbon dioxide at various concentrations, from industrial sources or from the air, can be performed using nanoporous adsorbent materials. However, the question of accurately identifying specific CO$_2$ sorption mechanisms in solids is an important step for materials design optimisation. Up to now, the use of first principles or \textit{ab initio} calculations to accurately describe molecular interactions in such systems often yields imprecise solutions\cite{odoh12,klimevs2012perspective}. Due to the natural way in which many-body interactions can be treated, as well as the sheer size of the computational space, quantum computing represents a future alternative in modelling such systems. Whilst contemporary quantum computing solvers are successful in capturing many-body interactions of a chemical system, the number of orbitals is limited such that usually only small molecules can be treated. In this work, we develop a strategy for accurately describing molecular interactions with quantum computers with a special emphasis on modelling CO$_2$ capture with Metal-Organic Frameworks (MOFs), a candidate for scalable carbon-capture technology. We anticipate that the insights obtained with our study can be used to feed empirical force field calculations. Here, the aluminium fumarate MOF is decomposed into interacting fragments, and the fragment containing the adsorbing Al site is treated with quantum computing. The interaction of CO$_2$ with the aluminium active site is modelled using the Variational Quantum Eigensolver (VQE) - a hybrid quantum classical algorithm. These results are compared to classical methods and highlight which quantum solver and fragmentations can be successful in capturing many-body interactions in MOF-CO$_2$ systems. The results obtained allow us to get an insight into the complex mechanism of guest-host bond formation.

Global warming can be considered as the greatest challenge of our century. A global energy transition based on low-carbon emissions is thus urgently needed to limit global warming below 2$^{\circ}$C in the next few decades. To tackle deleterious greenhouse gas effects and particularly CO$_2$ contributing to $\sim$70\% of the overall emissions \cite{pachauri2014}, drastic changes have to be made. This includes the swift introduction of policies and fundamental political changes in order to rapidly shift from the use of fossil fuels to low-carbon energy sources.

Along with avoiding fossil carbon and reducing its use, for example with renewable energy,  carbon capture and storage (CCS) is considered a complementary strategy to curb greenhouse gas emissions, as a robust means to target the decarbonation challenge \cite{tapia2018,lecomte2010}. In CCS, the first step is to capture the CO$_2$ either from anthropogenic point sources, from bioenergy conversion or from atmosphere by direct air capture to remove current and historical emissions \cite{gambhir2019,lehtveer2021}. CO$_2$ adsorption in solid porous sorbents such as carbon\cite{shen2013}, zeolites\cite{liu2012}, covalent-organic polymers\cite{wang2019}, covalent-organic frameworks \cite{gao2015} and metal-organic frameworks (MOFs)\cite{piscopo2020} has drawn widespread attention due to their low energy requirements \cite{li2017}. MOFs are nano- and/or mesoporous synthetic crystals, composed of metal ion/oxide nodes coordinated by organic ligands. They can be compared to molecular LEGO, offering a quasi-infinite tunability with respect to their pore sizes and reactivity depending on their metal, ligand type and overall chemistry\cite{collins2016}. MOFs have emerged as promising candidates for the carbon capture technology of the near future\cite{shekhah2011}, due to a unique versatility, enhanced by emerging machine learning approaches to rapidly propose structures\cite{wilmer2012large}.

Nevertheless, water, often present in industrial flue gases or in the air, can strongly affect CO$_2$ capacity and selectivity. 
To predict properties of next generation MOFs, high throughput techniques are usually employed, which consist of 
screening of a very large number of hypothetical MOF structures, followed by the determination of isotherms to find the best candidate for targeting applications\cite{dureckova2019}. In the latter, thermodynamic calculations are based on empirical force fields \cite{bureekaew2013,boyd2017}, adjusted from density functional theory (DFT) or \textit{ab initio} techniques, and then used in Monte Carlo algorithms. DFT has been successful in predicting a number of material properties, however, its application is based on approximate functionals, known to not fully capture electronic structure when van der Waals interactions are present \cite{klimevs2012perspective}, such as in the case of CO$_2$ capture on MOFs \cite{poloni2012co2, odoh12,vlaisavljevich2017performance}. In particular, MOF point charges participating in adsorbate-adsorbent interactions have been shown to strongly depend on the choice of DFT functional, basis set size, \textit{etc}\cite{vlaisavljevich2017performance, sladekova2020}.
\begin{figure*}[!h]
\centering
\includegraphics[width=0.8\textwidth]{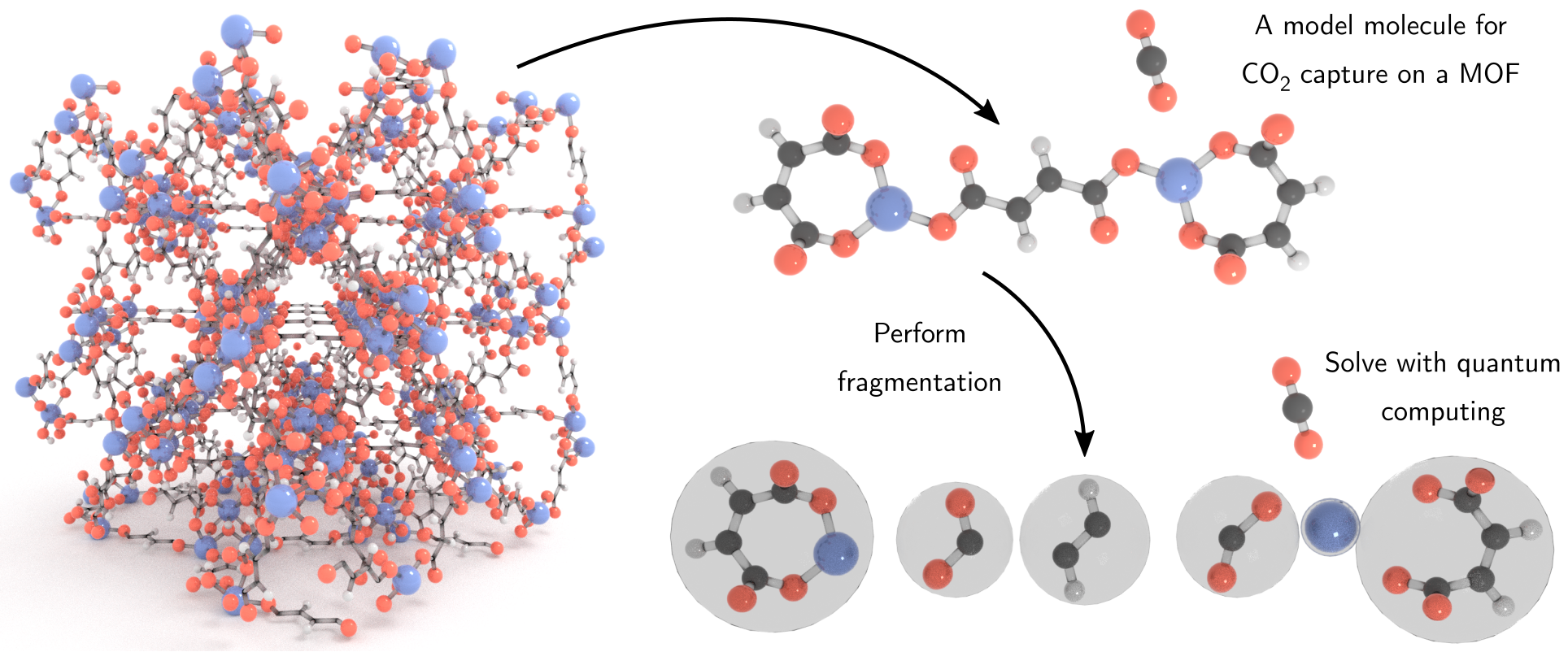}
\caption{\textbf{A quantum computing methodology applied to CO$_2$ capture on Metal-Organic Frameworks}. Al-fumarate molecule, a building block found in water stable MOF structures\cite{alvarez2015,bozbiyik2016,tannert2019,ke2018}, has been chosen to investigate CO$_2$ adsorption in Al based MOFs as as potential candidate for CO$_2$ capture \cite{gaab12,llewellyn_book_13} (left). After fragmenting the fumarate (right), binding of the active Al site with the CO$_2$ molecule is determined by quantum computing, while fumarate fragments (gray bubbles) are treated with a classical computer quantum chemistry solver.}
\label{fig:TOC}
\end{figure*}
\textit{Ab initio} algorithms, that can in some cases reach an exact solution of the Schr\"odinger equation, are time consuming and remain limited for many-body systems, where solutions in practice do not fully converge. Upscaling simulations based on such techniques yield in some cases spurious isotherm predictions, especially when accounting for ubiquitous moisture\cite{sladekova2020,jajko2021,ymagnin2022}, that cannot be neglected in a screening process because of its critical role in applications \cite{boyd2019}. 

Quantum computing is a promising tool for many-body systems, chemistry, materials sciences, \textit{etc}\cite{veiss_ch2,sugisaki19,ggd_mr_21,liu20, yoshioka20, manrique_21, yamamoto22}. Such advanced techniques, benefiting from lower levels of approximations could thus improve the quest in MOF design, with the high degree of accuracy required when such sorbents interact with rich adsorbates media. In quantum computers, qubit (quantum version of the classical bit) states can take an infinity of values between 0 and 1 due to superposition phenomena. In quantum simulations, qubit states are manipulated using quantum gates to mimic electronic wavefunctions, while entanglement phenomena allow for high computing efficiency. Such an approach is in principle ideal for working with electrons, since the exponentially large space of electronic states can be captured by the exponentially large Hilbert space of qubit states. This is a key advantage over classical simulations which are hindered by the exponential scaling of the electronic structure problem. However, current noisy intermediate scale quantum computing hardware (NISQ) is limited by the loss of quantum properties (quantum decoherence), due to undesired interactions of qubits with their environment, restricting first principles simulations to small and simple systems\cite{nisq_preskill}. 

\section*{Results and discussions} To overcome the limitations of current day quantum computing methods, a fragmentation technique based on Density Matrix Embedding Theory (DMET)\cite{knizia13, wouters16} has been proposed as a robust and versatile approach, where hybrid numerical methods mixing both quantum and classical solvers can be applied to the different pieces of a fragmented material, Fig.\ref{fig:TOC}. In DMET, low cost methods such as Hartree-Fock can be applied to non-active molecule fragments (\textit{i.e.} not participating in a physical or chemical reaction studied), while active fragments can be handled using state-of-the-art quantum computing. As depicted in Fig.\ref{fig:TOC}, CO$_2$ interaction with the Al-fumarate cluster is investigated to determine the adsorption energy associated with CO$_2$. An Al-fumarate molecule has one Al surrounded with oxygen atoms. In a realistic MIL-53(Al) two Al atoms would be relatively close by, even forming hydrogen bonds via their respective OH$^-$ groups. However, since quantum computers are limited in computational space and since studies show that only one of the two neighboring Al atoms participates in CO$_2$ capture \cite{serre2007explanation,damas2021understanding} we chose the Al-fumarate molecule as a model system. Our choice of an Al-based MOF as a center of the study is also motivated with the fact that the electronic structure of Al-fumarate molecule could be relatively tame. In such a case low level solvers such as restricted Hartre-Fock provide a reasonable estimation to the ground state energy and could be used as a benchmark of the methodology itself.

The explicit calculation of the adsorption energy from first principles is intractable for large systems using brute force classical techniques due to the exponential scaling of the electronic structure problem. As indicated above, quantum computing techniques could overcome this problem, however devices which are sufficiently free from error, or which exhibit gate stabilities that allow for error-correcting schemes are still needed. In anticipation of this "fault-tolerant" regime\cite{egan21}, and to pave the way for future quantum simulations of solid-gas interactions, we study physical adsorption by NISQ-era algorithms. To do so, we use fragmentation strategies where the preferential adsorption site is solved by simulated quantum computation. In the latter, the adsorbing Al atom is treated with unitary coupled cluster truncated to singles and doubles excitations (UCCSD)\cite{anand21}, an ansatz for the electronic wavefunction that can be variationally optimised by quantum computing techniques such as the variational quantum eigensolver (VQE) \cite{peruzzo}. 

VQE is a hybrid quantum-classical algorithm in which a quantum computer is used to store a quantum state, while measured values of its energy are fed to a classical computer which variationally optimises the energy. Technical details are provided in the Methods section. To limit the number of orbitals in the calculation (which reduces the number of qubits), some orbitals have been frozen out (electronic correlation not included), leaving an "active space" of orbitals contributing to the correlation, and in the context of the UCCSD ansatz we refer to this as AS-UCCSD. To compare the proposed approach to well established classical methods, fragments corresponding to low adsorption energy sites are also determined from classical solvers such as full space CCSD, mean-field restricted Hartree-Fock (RHF), and second-order M\o ller-Plesset perturbation theory (MP2), while fragments not directly involved in binding are limited to RHF (neglecting electronic correlation) or MP2 (a cheaper approximation to electronic correlation). The effects of different fragmentation scenarios with alternative solver mixing are presented and discussed in the Supplementary Information.\\

 \begin{figure}[!h]
	\centering
	\begin{minipage}{0.4\textwidth}
	\begin{overpic}[width=1.0\textwidth]{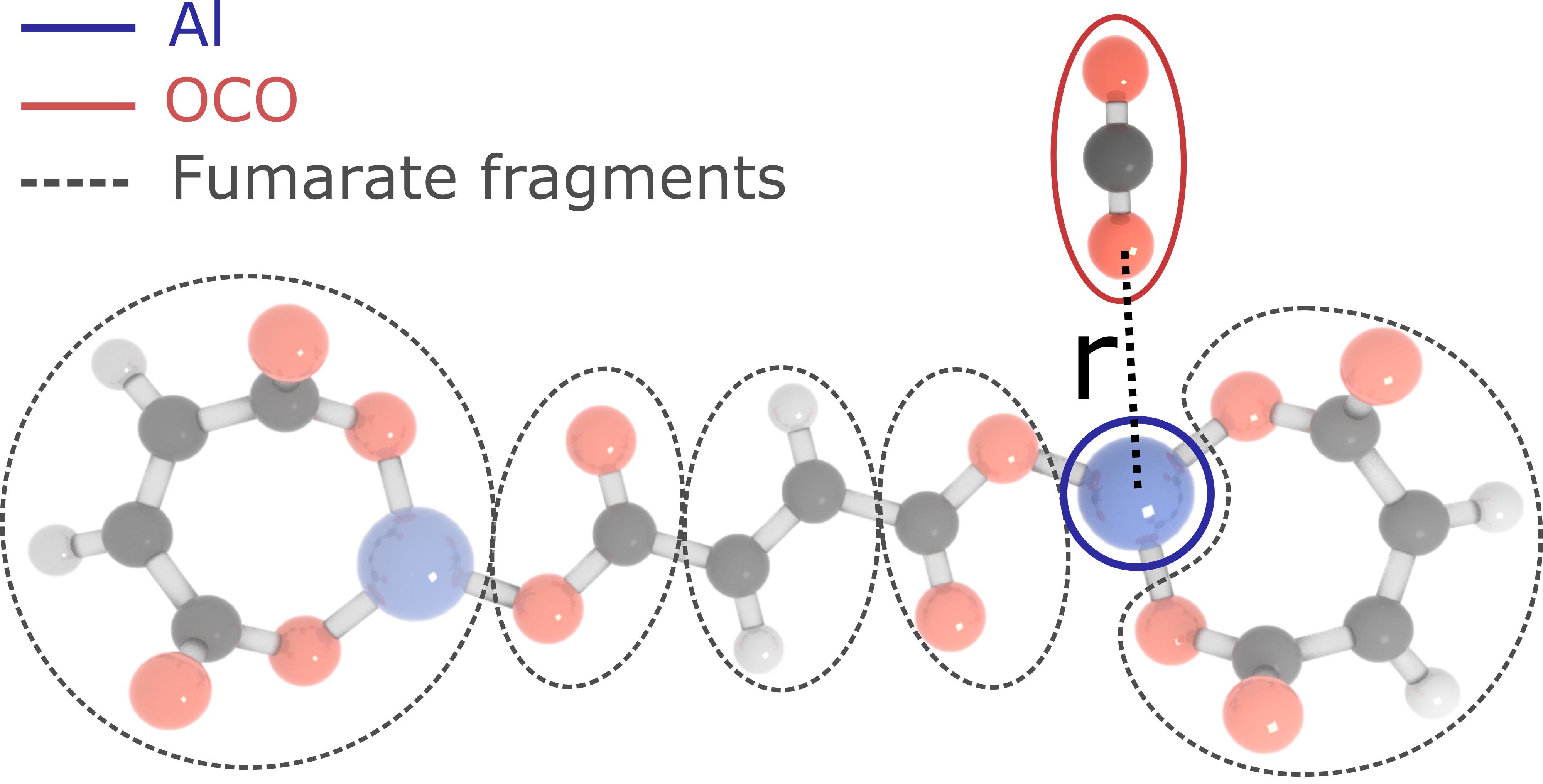}
	\put(-10,50){\textbf{a}}
	\end{overpic}
	\begin{overpic}[width=1.0\textwidth]{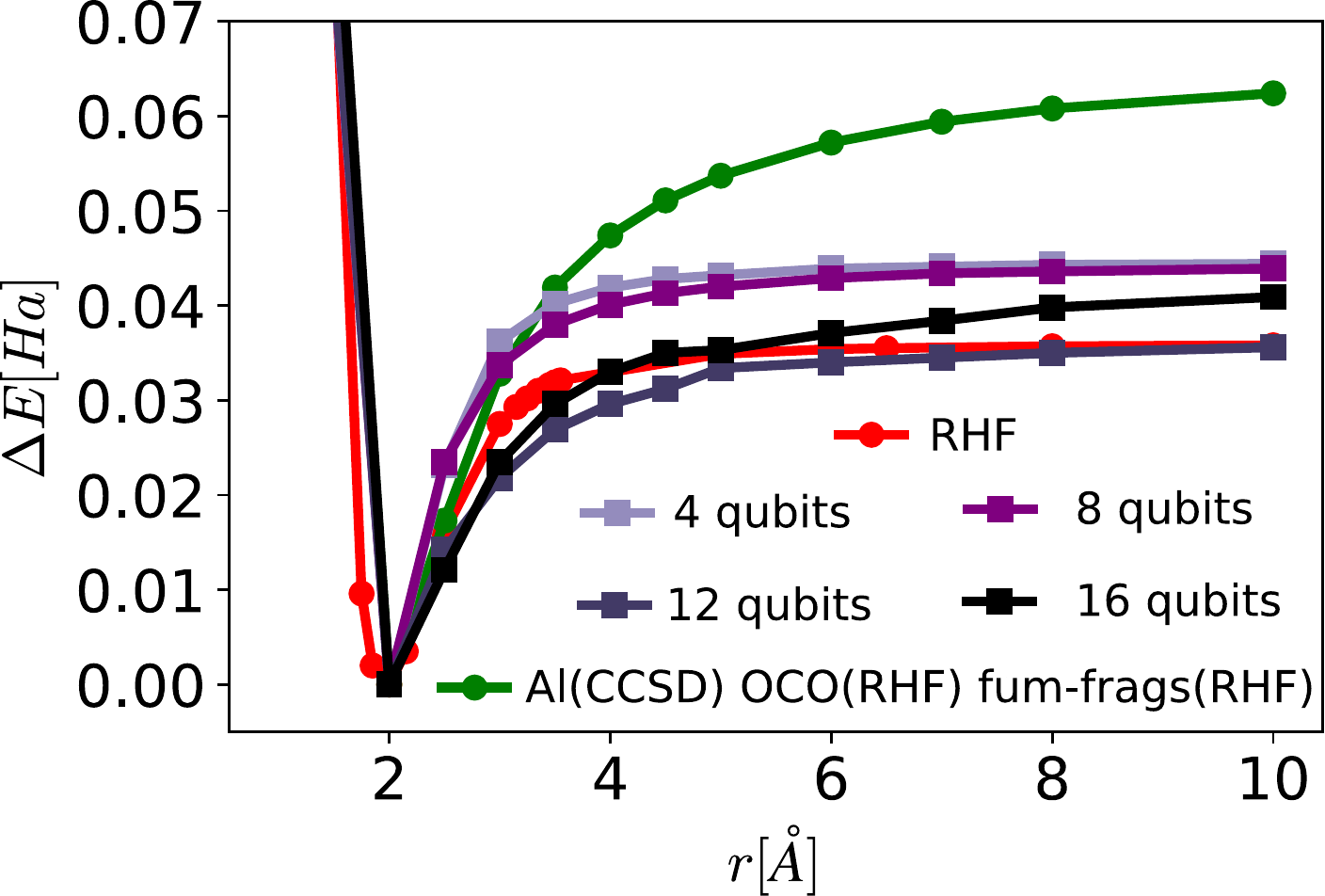}
	\put(-10,65){\textbf{c}}
	\end{overpic}
	\end{minipage}\hspace{1cm}\begin{minipage}{0.4\textwidth}
	\begin{overpic}[width=1.0\textwidth]{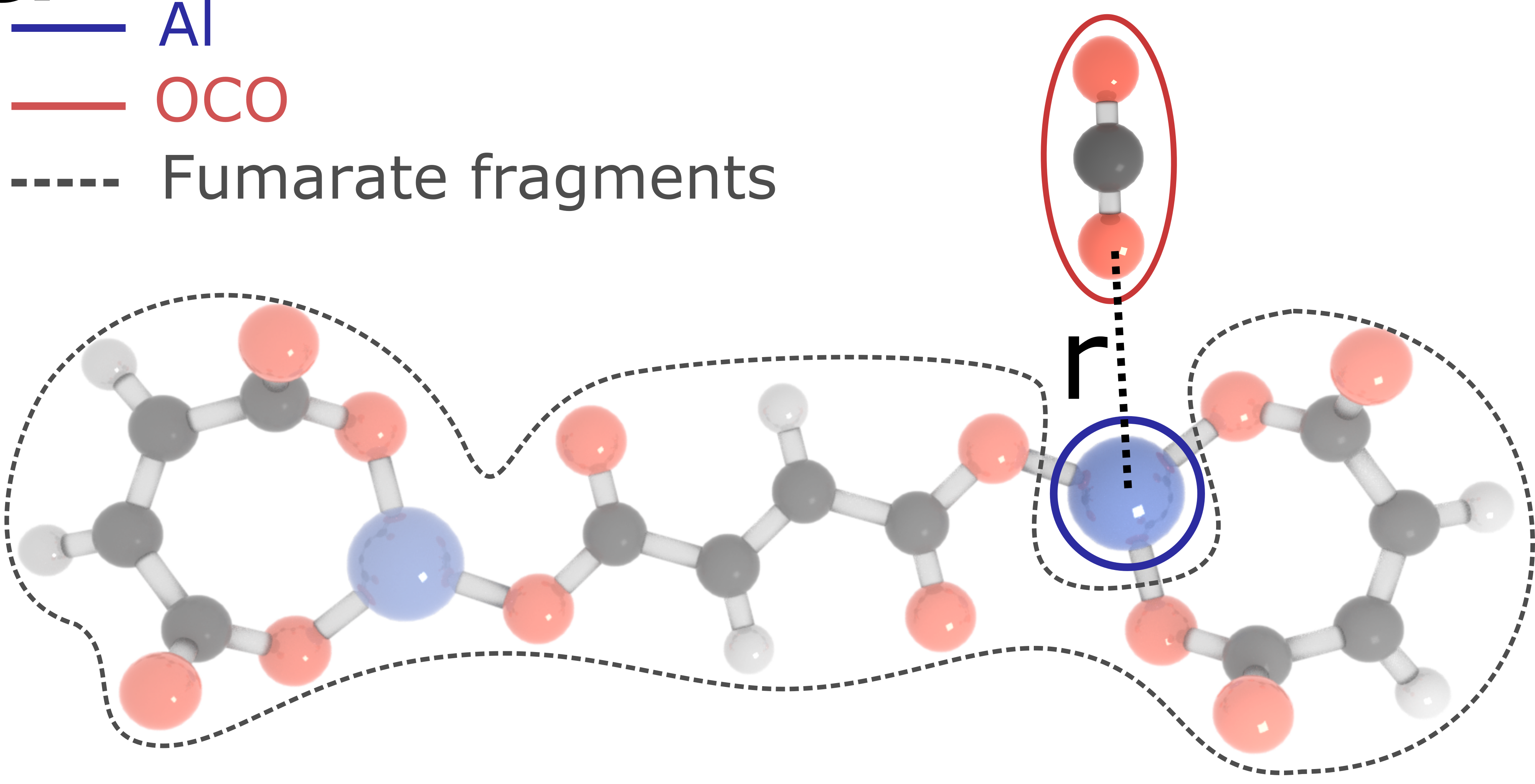}
		\put(-10,50){\textbf{b}}
	\end{overpic}
	\begin{overpic}[width=1.0\textwidth]{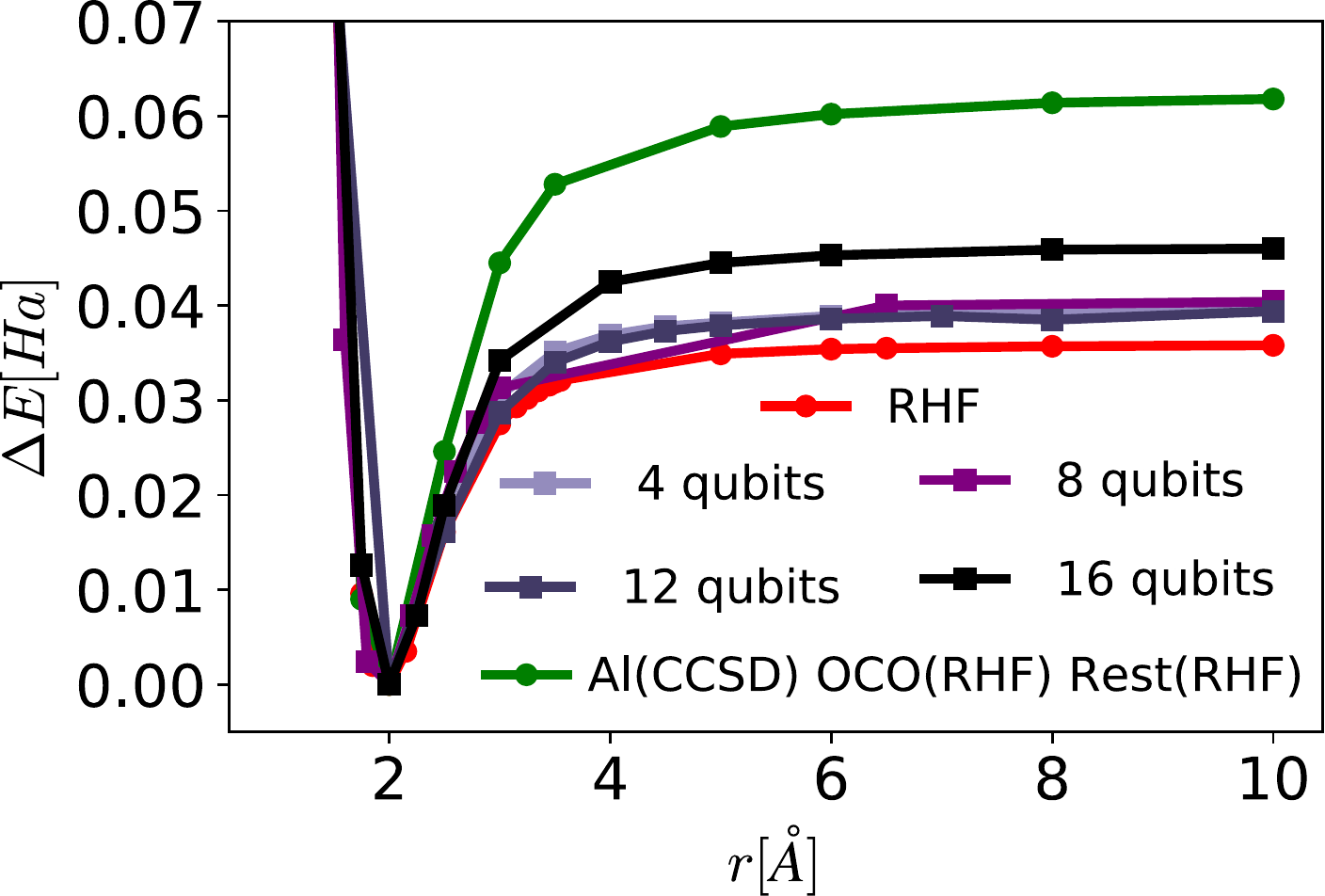}
	\put(-10,65){\textbf{d}}
	\end{overpic}
	\end{minipage}
	\caption{\textbf{Bond dissociation energy following different fragmentation schemes}. \textbf{a}, \textbf{b} Fragmentation strategies of the system with Al treated by quantum computing UCCSD or other solvers, CO$_2$ by RHF, and the ``rest'' of fumarate treated with RHF - fragmented ``rest'' of the fumarate corresponds to \textbf{a}, unfragmented fumarate to \textbf{b}. \textbf{c}, \textbf{d} Dissociation energy $\Delta E$ as a function of Al-CO$_2$ distance $r$ for the correlated fragment solver applied to the Al fragment with active space sizes corresponding to 4 to 16 qubits, compared to full space classical CCSD and RHF.}
\label{fig:Edis}
\end{figure}

In order to determine the Al-fumarate+CO$_2$ dissociation energy, we first approximate the minimum energy position of the CO$_2$ relative to the fumarate using classical methods (see Supplementary Information). The total energy of the CO$_2$ + fumarate is then calculated by high accuracy quantum computing as a function of the distance ($r$) between the CO$_2$ molecule and the Al site of the fumarate for the two fragmentation schemes. In the first, Fig. \ref{fig:Edis}a, Al is treated by AS-UCCSD, the whole CO$_2$ molecule by RHF, and the rest of the fumarate is fragmented and treated by RHF. In the second, Fig. \ref{fig:Edis}b, the Al site (AS-UCCSD) and the CO$_2$ molecule (RHF) are embedded in a large mean-field environment, where the fumarate is treated as a single fragment (RHF). In both cases, quantum calculations are performed for different active spaces of the high level fragment, Fig. \ref{fig:Edis}c,d, both showing a minimum energy at $\sim$2 \AA. In Fig. \ref{fig:Edis}c, the 4 qubit (1 HOMO 1 LUMO orbital) and 8 qubit (2 HOMO 2 LUMO orbitals) cases show larger dissociation energy than RHF, while an active space of 12 qubits (3 HOMO 3 LUMO orbitals) tends to lower the bond dissociation energy back to the RHF value, and 16 qubits (4 HOMO 4 LUMO orbitals) increases the dissociation energy ($\Delta E$($r$=10 \AA), which we refer to as $\Delta E$ from hereon in the text) to a value lying between the 8 qubit and RHF values. By comparison, the full-space CCSD solver for Al also exhibits a larger value than RHF, indicating that correlation contributes positively to $\Delta E$, while also showing the qualitative consistency between quantum computational and high level classical approaches.

Dissociation energies corresponding to the second fragmentation (Fig. \ref{fig:Edis}b), are shown in Fig. \ref{fig:Edis}d. Here, we observe that the 16 qubit case for AS-UCCSD applied to the Al fragment moves the dissociation energy (relative to smaller active spaces) towards the full space CCSD curve. In Tab. \ref{dmet_edis_vs_norbs_rest_al_oco}, we show the difference between the classical CCSD and the quantum UCCSD as a function of active space for fragmentation Fig. \ref{fig:Edis}b. Differences in $\Delta E$, for comparable active spaces, are found $<$ 1 mHa. We also note key differences between the two fragmentations. While for both cases $\Delta E$ is significantly lower at 16 qubits compared to CCSD, fragmentation \ref{fig:Edis}a shows a larger gap between AS-UCCSD and CCSD in $\Delta E$). Also, fragmentation \ref{fig:Edis}a exhibits smaller $\Delta E$ at 16 qubits than for 4 or 8 qubits, at variance to fragmentation \ref{fig:Edis}b. This (combined with the small differences in correlation energy for larger active spaces shown in Fig. \ref{dmet_ecorr_vs_norbs_restsplit_al_oco} and \ref{dmet_ecorr_vs_norbs_rest_al_oco}) suggest differences between these fragmentations in the treatment of correlation for UCCSD compared to CCSD, which may impact the long-range interactions between Al and CO$_2$. Comparing to results from the literature, we note that recent works have predicted binding energies of similar order of magnitudes from classical DFT calculations \cite{damas21}.

Note the sensitivity of DMET to fragmentation has been observed by our team for other molecules (not shown here), also revealing a strong qualitative dependence in $\Delta E$ depending on fragmentation schemes in those systems. It is also important to note that our results have strong implications for the notion of ``democratic'' mixing of local fragment properties (\textit{e.g.} the fragment 1-RDM and 2-RDM -reduced density matrices-), reported to be sub-optimal when different solvers are applied to different fragments \cite{wouters16}. Contrary to these works, we demonstrate that when correlated fragment solvers (\textit{i.e.} post-HF) are involved, physical dissociation curves are only found when different solvers are used for the fragment bonding to CO$_2$ and for the other fragments (Fig. \ref{fig:sm_edis_frags} in Supplementary Information).  
Therefore, we find that physically reasonable models of Al-fumarate+CO$_2$ dissociation can be obtained from DMET using democratic mixing of different solvers combined with the appropriate fragmentation. Overall, our results show that quantum computing methodologies can be used for studying these kinds of systems.
\begin{table}[h!]
	\begin{center}
		\begin{tabular}{ c|c|c|c|c|c|c|c|c|c|c|c} 
			
			\hline

			 Nr. of active orbitals  & 0 & 2 & 4 & 6 & 8 & 9 &10 & 12 &14 & 16 & 18\\
			\hline
			\hline
			 $\Delta E_{CCSD}[10^{-3}Ha]$ & 35.8 & 39.6 & 40.3 & 39.6 & 46.5 & 51.6 & 51.2 & 62.1 & 57.7 & 60.4 & 61.8\\

			 $\Delta E_{UCCSD}[10^{-3}Ha]$ & 35.8 & 39.4 & 40.4 & 39.4 & 46.0 & 51.3 & / & / & / & / & /\\
			\hline			
			 Difference  & 0 & 0.2 & 0.1 & 0.2 & 0.5 & 0.3 &/ & / &/ & / & /\\
			\hline
			\hline
		\end{tabular}
	\end{center}
	\caption{Dissociation energy ($\Delta E$($r$=10 \AA)) versus active space size for CCSD and UCCSD fragment solvers, using the fragmentation strategy depicted in Fig. \ref{fig:Edis}b. For even numbers of active orbitals, equal numbers of occupied and unoccupied orbitals are used. For the case when 9 active orbitals are considered, 4 of them are HOMO and 5 of them LUMO. \label{dmet_edis_vs_norbs_rest_al_oco}}
\end{table}	
%
%

\begin{figure*}[!h]
\centering
\includegraphics[width=1\textwidth]{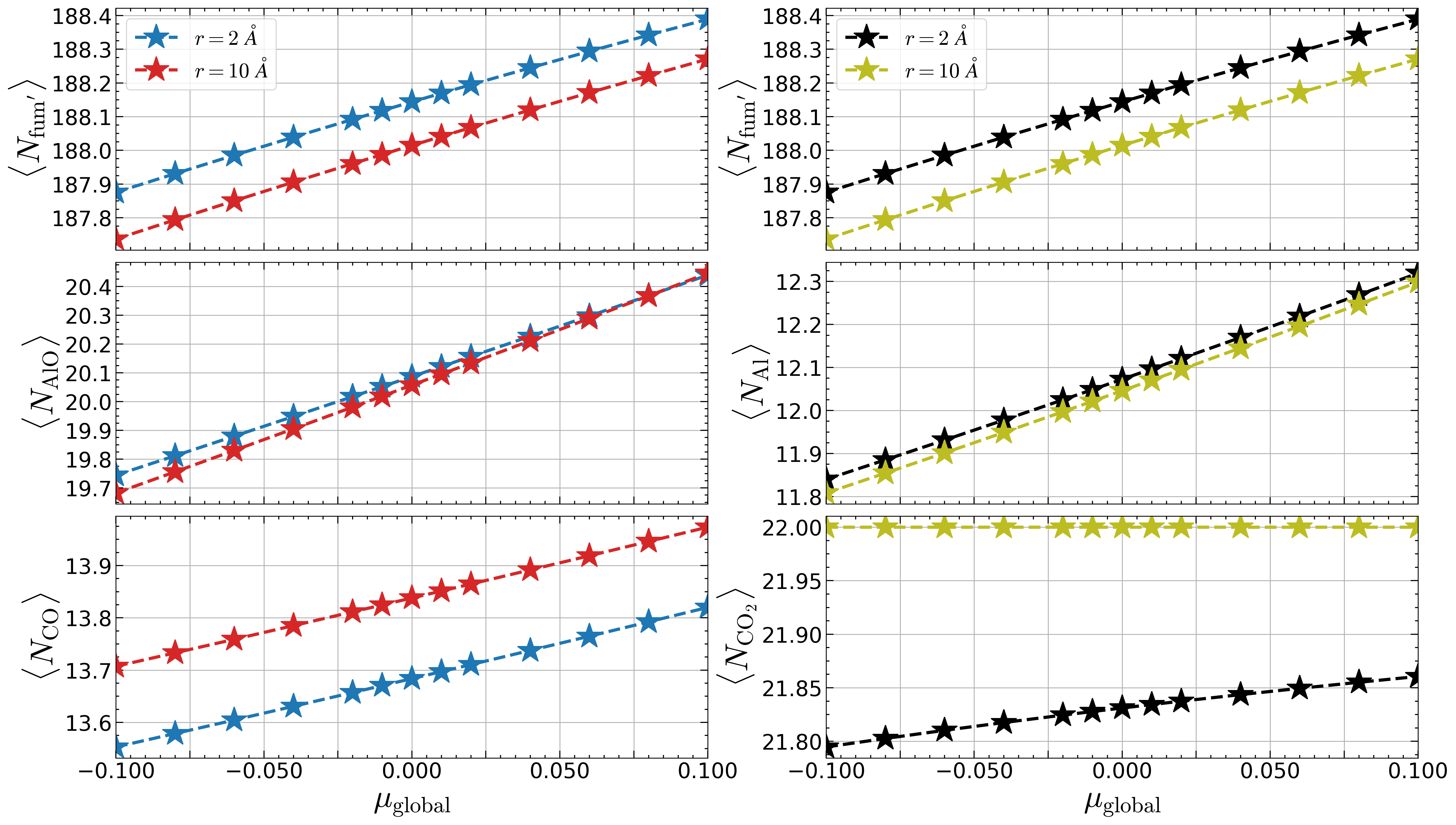}
\caption{\textbf{Local particle number versus global chemical potential}. For some fragmentations, charge transfer can occur between the Al-fumarate and CO$_2$ even at large distance. For other fragmentations this does not occur. Here fum' refers to the fumarate without Al atom.}
\label{fig:pnum_mu}
\end{figure*}

\begin{figure*}[!h]
\centering
\includegraphics[width=0.7\textwidth]{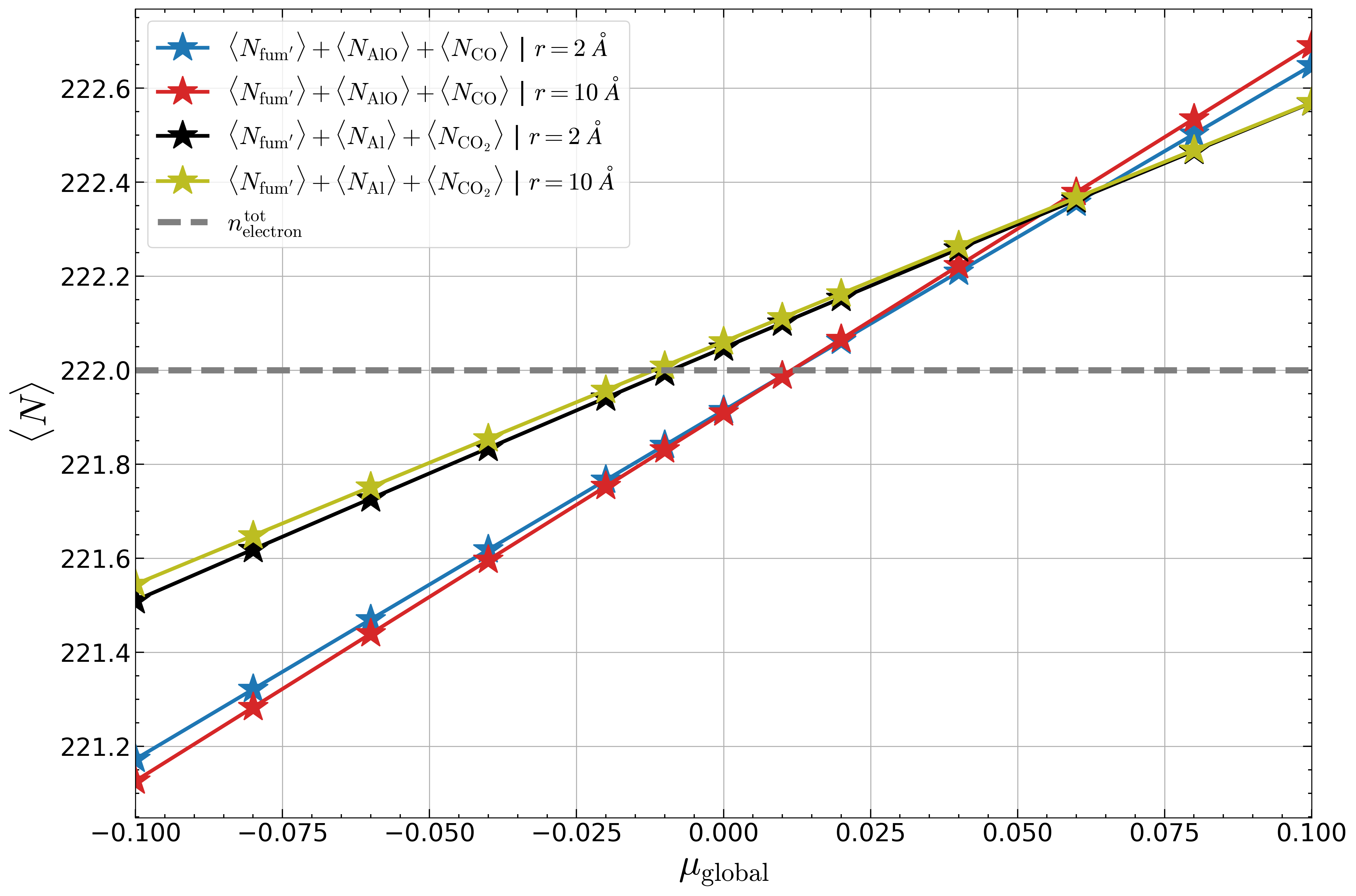}
\caption{\textbf{Total particle number versus global chemical potential}. In DMET, the global chemical potential is used to ensure the total number of electrons reproduces the trace of the total density matrix. At $\langle N \rangle$ = 222.0, $\mu_{\rm global}$ is approximately -0.01 (0.01) for the fragmentation shown in Fig. \ref{fig:Edis}b (Fig. \ref{fig:sm_edis_frags}d). Here fum' refers to the fumarate without Al atom. The total system contains 222 electrons and this is denoted with a dashed grey line.}
\label{fig:tnum_mu}
\end{figure*}

 To go a step further, we investigate consequences of using the 1-shot version of DMET, where a global chemical potential $\mu_{global}$ is optimised such that the sum of fragment electron numbers matches the total number of electrons of the system, with no other parameters in the cost function of the DMET algorithm (details provided in Supplementary Information). While such an approach leads to less variational flexibility, it benefits from higher efficiency and is commonly used in quantum computational applications \cite{li21, kawashima21, yamazaki18}. 
Since the 1-shot DMET algorithm attempts to optimise a single global chemical potential, it stands to reason that unphysical dissociation in Fig. \ref{fig:sm_edis_frags} may be related to the use of a single parameter to describe electron transfer between fragments. To investigate this hypothesis, we determined the local particle number of each fragment ($\langle N_x \rangle$ for fragment $x$) for a given $\mu_{global}$, then sum over $x$ to determine the total particle number ($\langle N \rangle$) of the system as a function of $\mu_{global}$, Fig. \ref{fig:pnum_mu} and \ref{fig:tnum_mu}. We then select the local particle number of each fragment for the chemical potential that conserves the total electron number ($\langle N \rangle$ = 222.0), Tab. \ref{dmet_n_versus_mu_alo_co_rest_vs_al_co2_rest_table}.

\begin{table}[h]
\centering
\begin{tabular}{|c||c|c|c|c|}
\hline
Fig. \ref{fig:sm_edis_frags}d fragmentation & $\langle N_{\rm CO}\rangle$ & $\langle N_{\rm AlO}\rangle$ &  $\langle N_{\rm fum'}\rangle$ & $\sum_{x}$  \\
\hline 
$r = 2 \text{ \AA}$ & $13.6972$ & $20.1216$ &  $188.1694$ & $221.9881$ \\
$r = 10 \text{ \AA}$ & $13.8514$ & $20.0954$ &  $188.0411$ & $221.9878$ \\
\hline
$\langle \delta N\rangle$ & $-0.1542$ & $0.0262$ & $0.1283$ & $0.0003$ \\
\hline
Fig. \ref{fig:Edis}b fragmentation & $\langle N_{\rm CO_2}\rangle$ & $\langle N_{\rm Al}\rangle$ & $\langle N_{\rm fum'}\rangle$ & $\sum_{x}$  \\
\hline 
$r = 2 \text{ \AA}$ & $21.8278$ & $12.0487$ & $188.1180$ & $221.9944$ \\
$r = 10 \text{ \AA}$ & $22.0000$ & $12.0215$ & $187.9876$ & $222.0091$ \\
\hline
$\langle \delta N\rangle$ & $-0.172186$ & $0.027151$ & $0.13035$ & $-0.0147$ \\
\hline
\end{tabular}
\captionof{table}{Local electron number for $\mu_{\rm global} \approx +0.01$ (fragmentation in Fig. \ref{fig:sm_edis_frags}d) and $\mu_{\rm global} \approx -0.01$ (fragmentation in Fig.\ref{fig:Edis}b). The rightmost column corresponds to a sum over fragments ($x$ represents a fragment label), for which the total number should sum to 222 and the total charge transfer ($\langle \delta N\rangle$) should sum to 0, hence we note an error in electron number of approximately $\pm 0.01$ due to lack to strict DMET cost optimisation.}
\label{dmet_n_versus_mu_alo_co_rest_vs_al_co2_rest_table}
\end{table}

In the Fig. \ref{fig:tnum_mu}, we show that the total electron number is conserved for different chemical potentials depending on the fragmentation. When plots for the same fragmentation cross, the corresponding $\mu_{global}$ yields the same total electron number for different Al-O$_{CO_2}$ distances, and these crossings occur at different points for different fragmentations, meaning that the relation between $\mu_{global}$ and $r$ is not necessarily unique and depends on fragmentation.  
Moreover, for the CO$_2$+Al fragmentation (Fig. \ref{fig:Edis}) at $r = 10 \ $\AA, the DMET chemical potential does not change $\langle N_{CO_2} \rangle$, corresponding to $22$ electrons (see Fig. \ref{fig:pnum_mu}). As expected, the CO$_2$ is too far from the fumarate for charge transfer to occur. However, for the CO+(Al-O$_{CO_2}$) fragmentation (Fig. \ref{fig:sm_edis_frags}a,d), a charge transfer occurs even for large distances between the fumarate and CO$_2$. In other words, fragmentations shown in Fig. \ref{fig:sm_edis_frags}a,d do not prevent charge transfer through the whole system for any $r$, while that should be prevented for large Al-O$_{CO_2}$ distances. This could explain why certain DMET fragmentations can fail.

For the successful fragmentation strategies which exhibit physical dissociation curves, we show in Tab. \ref{dmet_n_versus_mu_alo_co_rest_vs_al_co2_rest_table} that decreasing $r$ leads to a charge transfer from the CO$_2$ to the fumarate fragments not containing the adsorbing Al atom. As CO$_2$ moves from $10$ \AA \ to $2$ \AA, about $0.17$ of a charge is transferred to the fumarate, $\sim 0.03$ is localised on the Al atom, while the remaining charge is distributed to the rest of the fumarate (electron numbers are not strictly conserved to better than $\pm$ 0.01 in Tab. \ref{dmet_n_versus_mu_alo_co_rest_vs_al_co2_rest_table} since points on the $\langle N \rangle$ vs. $\mu_{global}$ line do not correspond to fully optimised DMET cost functions). As expected, this further confirms that the Al atom corresponds to the most favorable adsorption site. Thus, most of the electronic charge does not localise on the Al site in the adsorption process, but is found to be redistributed between the CO$_2$ and the rest of the fumarate. Thus, the successful DMET fragmentation strategy shows that a complex bond spanning many atoms is formed with Al acting as an intermediary conducting site.

 \begin{figure}[!h]
	\centering
	\begin{minipage}{0.49\textwidth}
	\begin{overpic}[width=1.0\textwidth]{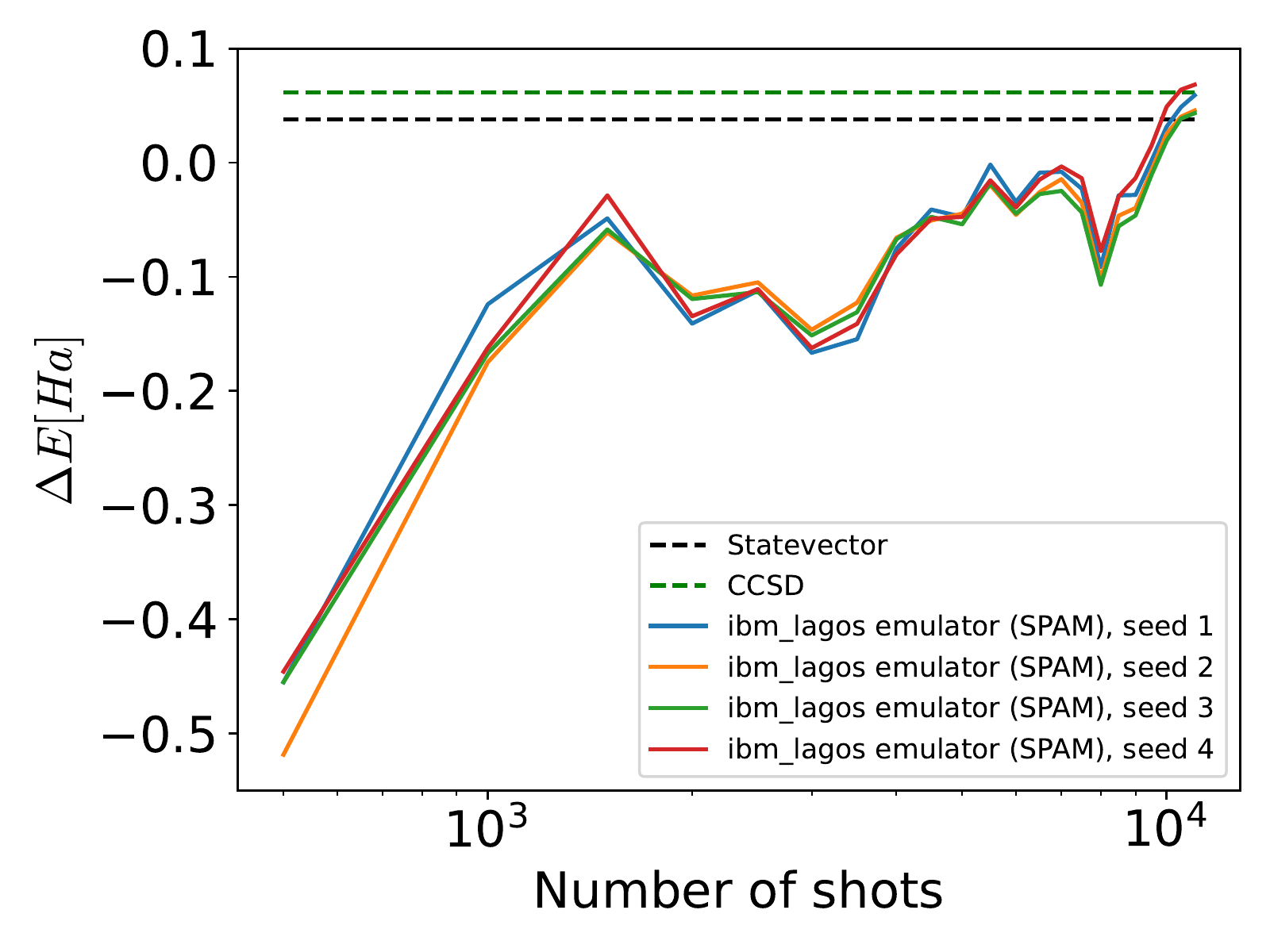}
	\put(-0.5,65){\textbf{a}}
	\end{overpic}
	\end{minipage}\hspace{0.25cm}\begin{minipage}{0.49\textwidth}
	\begin{overpic}[width=1.0\textwidth]{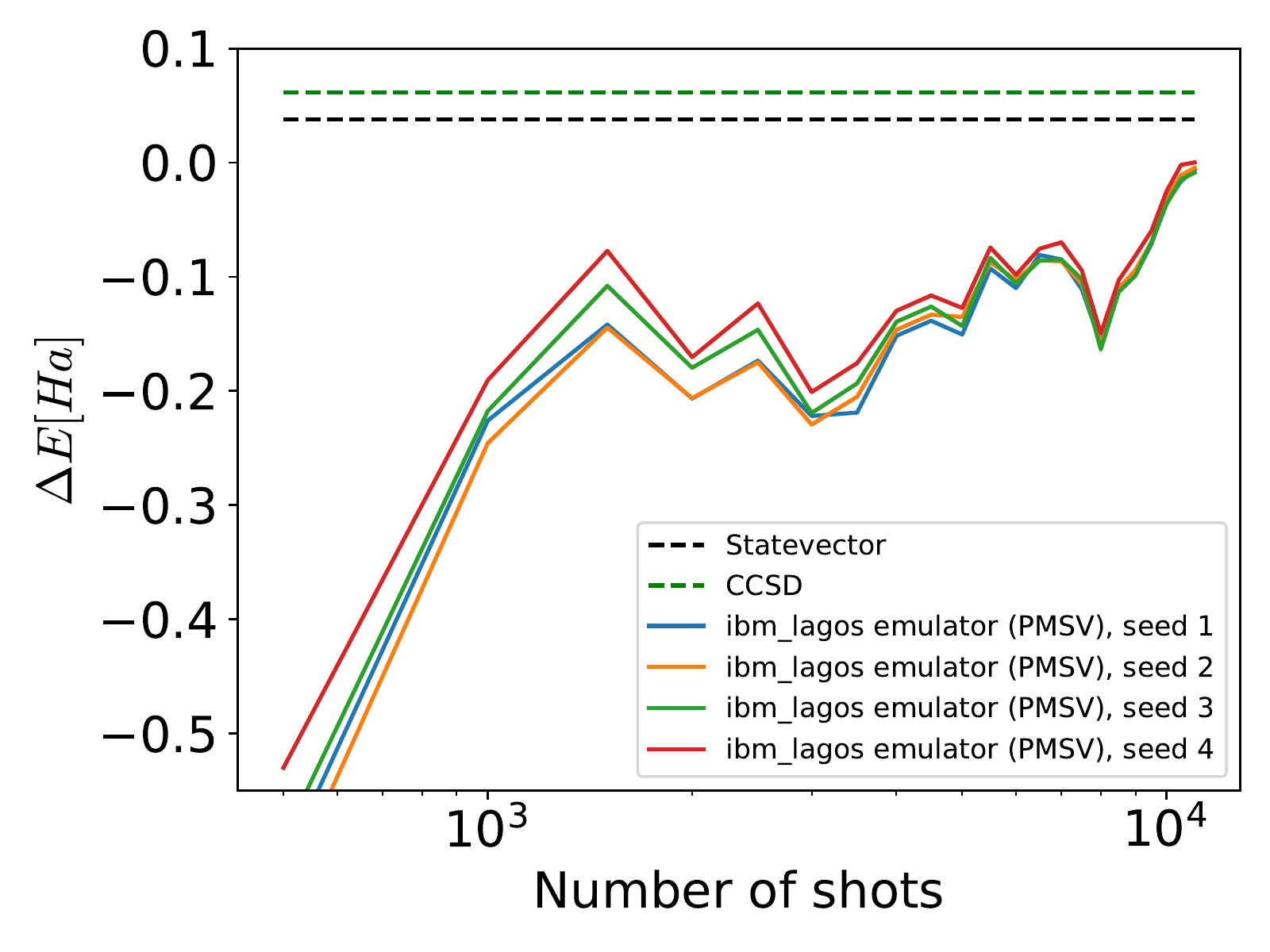}
	\put(-0.5,65){\textbf{b}}
	\end{overpic}
	\end{minipage}
	\caption{\textbf{Bond dissociation energy from noise-prone measurements of emulated hardware, using two error mitigation schemes}. \textbf{a} state preparation and measurement (SPAM) error mitigation. \textbf{b} post measurement symmetry verification (PMSV). In both cases, the same 4 randomisation seeds are used, and plotted as a function of the number of measurement shots. We note that SPAM mitigation and PMSV can lead to different results when calculating $\Delta E$. The fragmentation corresponds to that of Fig. \ref{fig:Edis}b}.
\label{fig:Edis_noise}
\end{figure}

Using the fragmentation shown in Fig. \ref{fig:Edis}b, we also present dissociation barrier calculations resulting from emulated hardware measurements obtained with typical hardware noise on an IBMQ device. Error mitigation schemes are currently necessary to minimise the effects of hardware noise. However, the impact on dissociation energy of the choice of error mitigation scheme is not well known. We report the dissociation energy ($\Delta E$) resulting from noisy measurements after applying one of two popular error mitigation schemes: mitigation of state preparation and measurement (SPAM) errors \cite{jackson15}, and partition measurement symmetry verification (PMSV) \cite{kirsopp22, yamamoto22}. Due to the stochastic nature of the measurement process, we plot the measured $\Delta E$ versus number of measurement shots, where a larger number of shots corresponds to a better representation of the statistical distribution of measurement results. This is repeated using 4 randomisation seeds for both error mitigation schemes applied individually for the same calibration data, which provides a systematic comparison between the schemes. The results are plotted in Fig. \ref{fig:Edis_noise}.

It can be seen that for a sufficiently small number of measurement shots, the resulting $\Delta E$ can be negative (Fig. \ref{fig:Edis_noise}a,b). We also note that the measured total energies for both geometries (not shown for brevity) are below the ideal noise-free value. Thus a negative $\Delta E$ implies that device errors for this noise model have a slightly greater impact on the dissociated geometry. In addition, we observe that $\Delta E$ approaches the large-shot limit differently for SPAM error mitigation and for PMSV. Hence, to calculate energy barriers using error mitigated measurements on quantum hardware, it should be noted that different geometries (which correspond to different Hamiltonian parameters and different ansatz parameters) can lead to different behaviour of a given error mitigation scheme, and these differences can accumulate leading to additional errors in the calculated energy barrier.

\section*{Conclusions} \label{discon}

 The work presented here demonstrates the application of quantum computing methods as high-accuracy post-Hartree-Fock solvers in the treatment of carbon capture on Metal-Organic Frameworks (MOFs). Our findings suggest that quantum computing methodologies are successful in capturing many-body correlations in the MOF+CO$_2$ system, with physical dissociation curves. Furthermore, the versatility of the embedded quantum computational approach allows for estimates of the bond dissociation energy of the van der Waals molecule formed between CO$_2$ and the unit cell of the MOF. Finally, sampling the expectation value of the number operator on different fragments gives an insight into the complex bond which the CO$_2$ forms with the MOF. Although the CO$_2$ binds to the Al-site, a charge transfer between CO$_2$ and the rest of the fumarate molecule occurs, while the number of electrons on the Al-site remains largely unchanged. This indicates a non-trivial dependence of the CO$_2$ binding energy on MOFs - the energy depends both on the metal which is the active site but also on the atoms surrounding the metal. In addition, we provide results from emulated hardware experiments with two popular error mitigation schemes for the purpose of calculating chemical dissociation barriers, and we found significant differences between SPAM mitigation and PMSV, which relate to differences in how the mitigation schemes treat the different geometries. In summary, this work opens the pathway towards the use of quantum computing for complex materials design with strong molecular interactions in view of real world applications such as greenhouse gas capture.

\section*{Methods} \label{methods}

In our calculations, we adopt a local cluster model of the full MOF supercell corresponding to the Al-fumarate molecule shown in Fig. \ref{fig:TOC}, whose metal ion/oxide nodes correspond to low energy sites where the sorption usually occurs. Given that quantum computing calculations of complex systems are still relatively expensive, in terms of circuit depth and the number of qubits, calculations are made with the STO-3G minimal basis set. We note that while minimal basis sets are insufficient to capture all many-body interactions and hence prevent an accurate prediction of the binding energy, we use minimal basis sets in order to qualitatively compare the performance of various post-HF methods and quantum computing methods as DMET solvers for this system.

In order to assess the minimum energy geometry of the combined Al-fumarate+CO$_2$ complex, geometry optimisation could be performed. However, long range dispersion interactions are not accurately captured at this level of theory. In particular, parameterised approximations to the dispersion interactions (such as DFT-D3 of Grimme et al.\cite{grimme2010}) are not reliable when using a minimal basis set. This is especially true for a MOF interacting with  CO$_2$ due to the complicated form of the guest-host exchange interaction, which DFT commonly fails to capture \cite{poloni2012co2}. Hence, to assess the optimal geometry, we carry out simple tests of the Al-fumarate+CO$_2$ bonding configurations with classical computing methodologies, for which both the fumarate and the CO$_2$ are kept fixed at their isolated ground state geometries. Classical computing calculations with (coupled cluster singles doubles (CCSD)\cite{bartlett_cc}) and without (Hartree-Fock (HF)) correlation of the combined Al-fumarate+CO$_2$ system were carried out for this purpose. These classical results serve as a guide for the combined quantum/classical simulations of the Al-fumarate+CO$_2$ system, in which the lowest energy bonding geometries (relative orientation, distance, incidence angle) can be assessed before tackling the problem with a quantum computational approach. From the results of the classical calculations, we select the bonding geometry of the Al-fumarate+CO$_2$ system to study using DMET with a quantum computational solver for the high accuracy fragment. Details and results of these classical calculations are reported in the Supplementary Information.\\
All quantum calculations in this paper are performed using Quantinuum's computational chemistry platform \cite{quantinuum}. This is a Python package for running quantum chemistry simulations on quantum computational hardware, built on top of the architecture agnostic quantum software compiler $\text{t}|\text{ket}\rangle^\text{TM}$ \cite{tket20, tket_url}. Quantinuum's computational chemistry platform utilises the classical chemistry package PySCF \cite{pyscf20} to generate classical data such as electronic integrals in the atomic orbital basis, in addition to classical methods such as HF, CCSD, and second-order M\o ller-Plesset perturbation theory (MP2)\cite{szabo_ostlund}. For HF calculations, we use the spin-restricted form (RHF) suitable for closed-shell systems. While all individual constituents of the Al-fumarate+CO$_2$ complex do not necessarily form a closed shell system, the DMET procedure itself generates a bath for each fragment, and each fragment+bath forms a closed shell system.
The mapping of our quantum algorithms to quantum circuits is performed using $\text{t}|\text{ket}\rangle^\text{TM}$ \cite{tket20} and Microsoft Azure is used as a platform for simulating idealised noise-free quantum hardware. For noisy simulations, we use the emulator of the \textit{ibm\_lagos} machine with a calibrated noise model, which is accessed via the IBMQ cloud service. This noise model includes qubit readout errors ranging from 5.1E-3 to 2.48E-2, single qubit Pauli-X errors ranging from 1.868E-4 to 3.025E-4, and CNOT errors ranging from 4.587E-3 to 1.037E-2. The full calibration data used for the simulation of this device is available from the authors on request. Two different error mitigation schemes are applied separately: state preparation and measurement (SPAM)\cite{jackson15} error mitigation, and partition measurement symmetry verification (PMSV)\cite{yamamoto22, kirsopp22}. For PMSV, we filter measurement shots for the $U(1)$ particle number symmetry. For more details on PMSV, we refer the reader to Yamamoto \textit{et al.}\cite{yamamoto22}

Density Matrix Embedding Theory (DMET) \cite{knizia13, wouters16} is an embedding procedure based on the Schmidt decomposition of the wavefunction into fragments and their complementary parts, which self-consistently combines correlated solutions across the whole system. A molecule is partitioned into a series of fragments, each one coupled to a bath that simulates the effect of the rest of the molecule. These reduced-size systems are then solved with different methods, with the most accurate method applied to the fragment representing the active site. More details about DMET can be found in Supplementary Information.

There are many possible ways to fragment the Al-fumarate+CO$_2$ complex. We report four fragmentation strategies found to show drastically different results, and investigate these differences. By active site we refer to that part of the fumarate involved in CO$_2$ adsorption. For solving non-active sites of the fumarate we rely on classical computing methods such as HF, MP2 or coupled cluster theories (see Supplementary Information for more details). For solving the active site, we compare classical coupled cluster theory with a quantum Unitary Coupled Cluster ansatz with single and double excitations (UCCSD) implemented with a Variational Quantum Eigensolver (VQE) \cite{peruzzo}. For the UCCSD fragment solver, orbital freezing was used to reduce the number of qubits required in the calculation and thus select an active space (AS-UCCSD). To investigate the effect of reducing the active space to feasible sizes, we plot the correlation energy as a function of active space size for all fragmentation strategies. The results, presented in the Supplementary Information, show that a significant degree of correlation is captured by the active spaces used in this work.

VQE is a hybrid quantum-classical algorithm which relies on a quantum computer to estimate the expectation value of energy, while relying on a classical optimiser to suggest improvements of the ansatz \cite{peruzzo}. In this work, we investigated the quantum UCC ansatz which has a complex circuit but is also (in principle) able to estimate the ground state energy of the system with higher precision. Thus, our strategy relies on UCC to obtain the ground state energy curve with the highest precision (see details in Supplementary Information). For noise-prone simulations corresponding to the \textit{ibm\_lagos} emulator, we take the converged VQE parameters of the state vector (idealised) DMET calculation, and re-calculate the active fragment energy using the hardware emulator with a calibrated noise model. 

A key quantity estimated is the CO$_2$-MOF bond stretching energy, $\Delta E(r)=E(r)-E$($2\,$\AA), which for $r\gg 2\,$\AA \ corresponds to the bond dissociation energy. This is justified as the CCSD numerical testing we performed showed that at $10\,$\AA \ distance the energy gradient is less than 0.5 mHa / \AA, hence the sum of the dissociated components is sufficiently well approximated by $E$($10\,$\AA). All energies are represented with respect to $E$($2\,$\AA), as classical calculations show that this is the bond distance between Al-fumarate and CO$_2$ which corresponds to the energy minimum (see Supplementary Information for details).

\section*{Declarations}
\subsection*{Ethical Approval and Consent to participate}
Not applicable.

\subsection*{Consent for publication}
Not applicable.

\subsection*{Data availability statement}
The authors will make all data available upon reasonable requests.

\subsection*{Competing interests}
The authors declare no competing interests.

\subsection*{Funding}
Marko J. Ran\v{c}i\'{c} and Wassil Sennane acknowledge the EU within the H2020 project $\langle$NE|AS|QC$\rangle$, Grant agreement ID: 951821 for funding.

\subsection*{Author contributions}

G.G.D. and D.Z.M directly obtained the results presented in the study. G.G.D., D.Z.M., W.S., M.C., M.J.R., and D.M.R interpreted the data. E.S., P.L., P.C. M.J.R. and D.M.R. conceptualised the study. G.G.D., D.Z.M, Y.M., P.L., M.C. M.J.R and D.M.R. wrote and organised the paper. P.C, M.K., M.J.R. and D.M.R supervised parts of the conducted work. All coauthors participated in joint discussions.

\subsection*{Acknowledgments}
Not applicable.

\bibliography{References}

\vfill
\pagebreak
\begin{center}
	\textbf{\huge Supplementary Information: Modelling Carbon Capture on Metal-Organic Frameworks with Quantum Computing}
\end{center}

\setcounter{equation}{0}
\setcounter{section}{0}
\setcounter{figure}{0}
\setcounter{table}{0}
\setcounter{page}{1}
\makeatletter
\renewcommand{\theequation}{S\arabic{equation}}
\renewcommand{\thefigure}{S\arabic{figure}}
\renewcommand{\thesection}{S\arabic{section}}

\section{Dissociation curves using different fragmentation strategies} \label{sec:sm_edis}

\begin{figure}[!h]
\centering
\includegraphics[width=13cm]{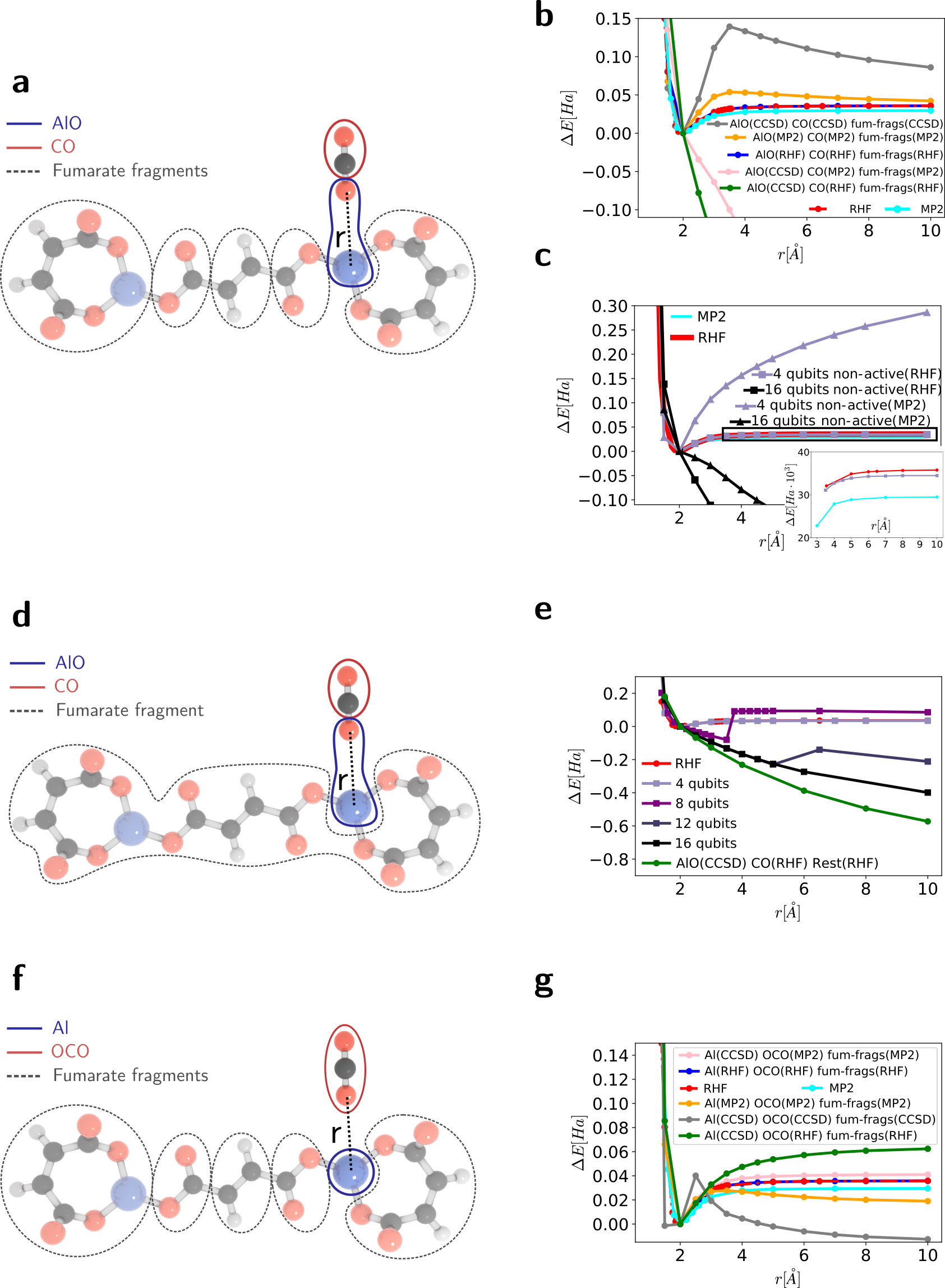}
\caption{\textbf{a.} Fragmentation strategy consisting of the fumarate Al and the O of CO$_2$ (O$_{CO_2}$) as one fragment (AlO), the CO of CO$_2$ as a separate fragment, and the rest of the fumarate divided into fragments. \textbf{b.} and \textbf{c.} show $\Delta E$ curves for fragmentation in \textbf{a} obtained using classical and quantum solvers, respectively. In \textbf{c} quantum AS-UCCSD is applied to the AlO fragment, while RHF or MP2 is used for all other fragments. The inset of \textbf{c} shows RHF and MP2 calculations (brute force, no DMET) compared to 4-qubit AS-UCCSD. \textbf{d.} Same as \textbf{a} but the fumarate (not including the Al bonding to CO$_2$) is one large fragment. \textbf{e.} shows $\Delta E$ curves for fragmentation in \textbf{d} obtained using classical and quantum solvers. \textbf{f.} Same fragmentation strategy as shown in Fig. \ref{fig:Edis}a, with $\Delta E$ curves shown in \textbf{g} for classical solvers. RHF and MP2 solvers for the AlO (\textbf{b}) and Al (\textbf{g}) fragments are also shown for comparison.}
\label{fig:sm_edis_frags}
\end{figure} 

The fragmentation strategies displayed in Fig. \ref{fig:sm_edis_frags} were also used to determine the Al-fumarate+CO$_2$ dissociation energy. Fig. \ref{fig:sm_edis_frags}a shows a fragmentation in which the stretched Al-O$_{CO_2}$ bond is contained in one fragment. When a correlated wavefunction solver CCSD (or AS-UCCSD with large enough active space) is used for a fragment containing the Al atom and the oxygen of CO$_2$ (the ``AlO'' fragment), with RHF or MP2 applied to the remaining parts of the system, the Al-fumarate+CO$_2$ complex is found to be unbound, with a monotonic decrease of the energy spanning all adsorbate-adsorbent distances $r$, (Fig. \ref{fig:sm_edis_frags}b,c). Such a behavior is inconsistent with brute force classical calculations, and with expected adsorption properties in MOFs \cite{gaab12, llewellyn_book_13}. Fig. \ref{fig:sm_edis_frags}d also shows a fragmentation consisting of the AlO fragment, but the rest of the fumarate is considered as one large fragment. The corresponding results in Fig. \ref{fig:sm_edis_frags}e show a similar conclusion - Al-fumarate+CO$_2$ complex is unbound for large enough active spaces of the AlO fragment. Hence, for these particular fragmentations, DMET is found to fail in describing the CO$_2$ binding to the Al site of the fumarate. 

In Fig. \ref{fig:sm_edis_frags}c,e we also show that the unphysical behaviour is manifested when a sufficiently large amount of correlation is included by sufficiently large active spaces. Indeed, the 4 qubit AS-UCCSD simulation (with RHF applied to the non-active fragments, \textit{i.e.} those fragments not directly involved in the Al-O$_{CO_2}$ bond) shows a bound state (purple squares), while the 16 qubit counterpart (black squares and black triangles) shows an unbounded state. Spurious dissociation behaviour can also be obtained when MP2 is used as a fragment solver, and the inset of Fig. \ref{fig:sm_edis_frags}c shows this is not exclusively due to the known weaknesses\cite{dutta2003full} of MP2 for dissociation in some systems.

Fig. \ref{fig:sm_edis_frags}f shows the same fragmentation as Fig. \ref{fig:Edis}a. Hence the results shown in Fig. \ref{fig:sm_edis_frags}g are the classical counterparts to Fig. \ref{fig:Edis}c. For this case, physical bound state dissociation curves without artifacts are observed, yet only when a mixture of solvers are used. When all fragments are solved with the same post-Hartree-Fock solver (MP2 or CCSD), discontinuities and/or non-monotonic behaviour is observed. Hence, the use of different solvers (maintaining democratic mixing), along with a careful choice of fragmentation appears to ameliorate the unphysical dissociation behaviour.

To summarise these results, no combination of DMET solver methods was found that resulted in qualitatively correct dissociation curves for larger active spaces when the high level fragment contains the stretched AlO bond, while in Fig. \ref{fig:sm_edis_frags}g it is shown that physical dissociation curves are not possible when the same solvers are used on all fragments (where the stretched bond lies between fragments). By comparing Fig. \ref{fig:Edis} to Fig. \ref{fig:sm_edis_frags}, our results show that small changes in fragmentation lead to large qualitative differences in dissociation behaviour - this is observed for small changes in fragmentation geometry and solver methods. 

\section{Orbital contribution to DMET correlation energy} \label{sec:orb_comb}

\begin{figure}[!h]
\centering
\includegraphics[width=9cm]{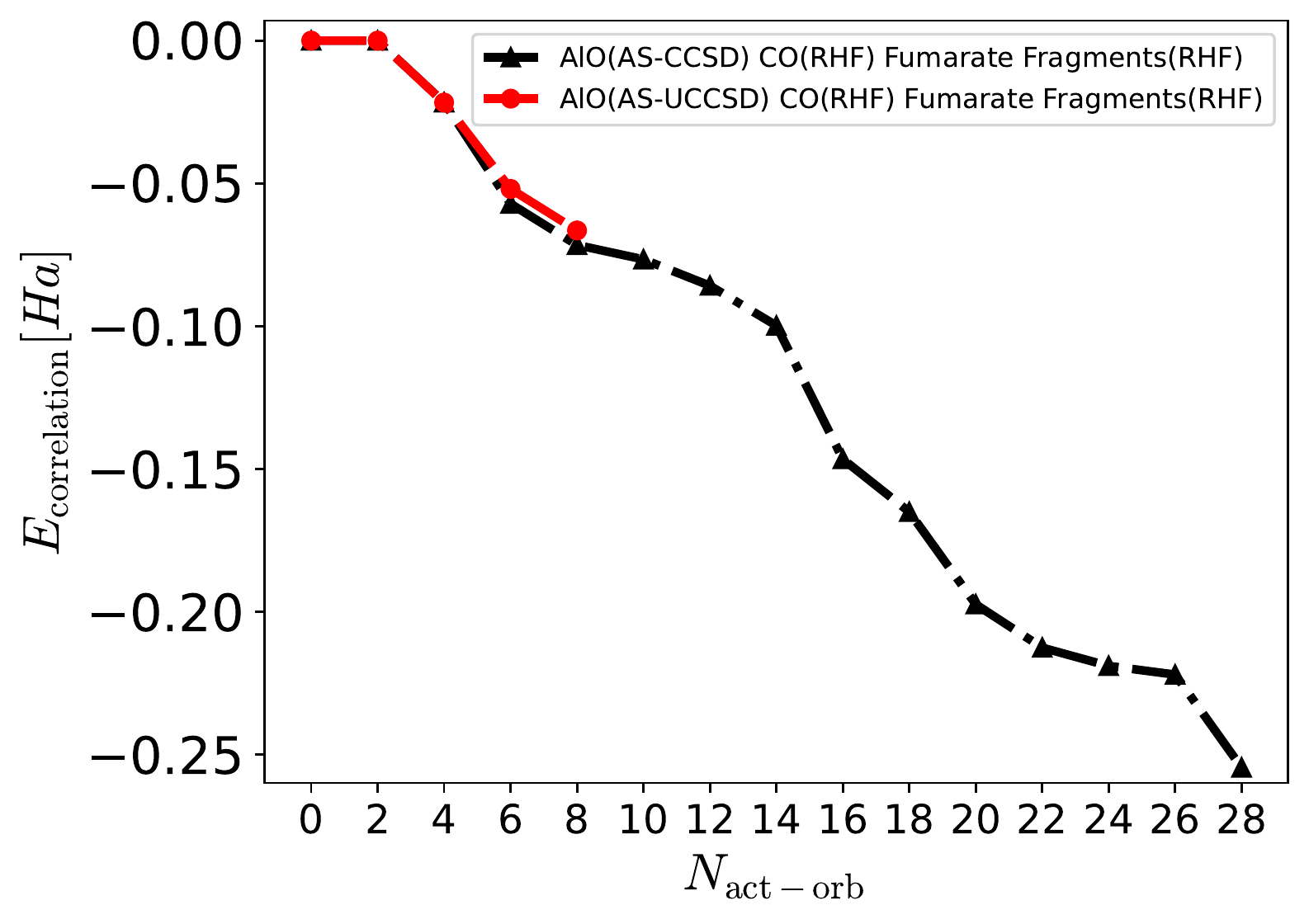}
\caption{Correlation energy versus active space size for CCSD and UCCSD fragment solvers, using the fragmentation depicted in Fig.\ref{fig:sm_edis_frags}a. $N_{\rm act-orb}$ refers to the number of active molecular spatial orbitals.}
\label{dmet_ecorr_vs_norbs_restsplit_alo_co}
\end{figure} 
\begin{figure}[!h]
\centering
\includegraphics[width=9cm]{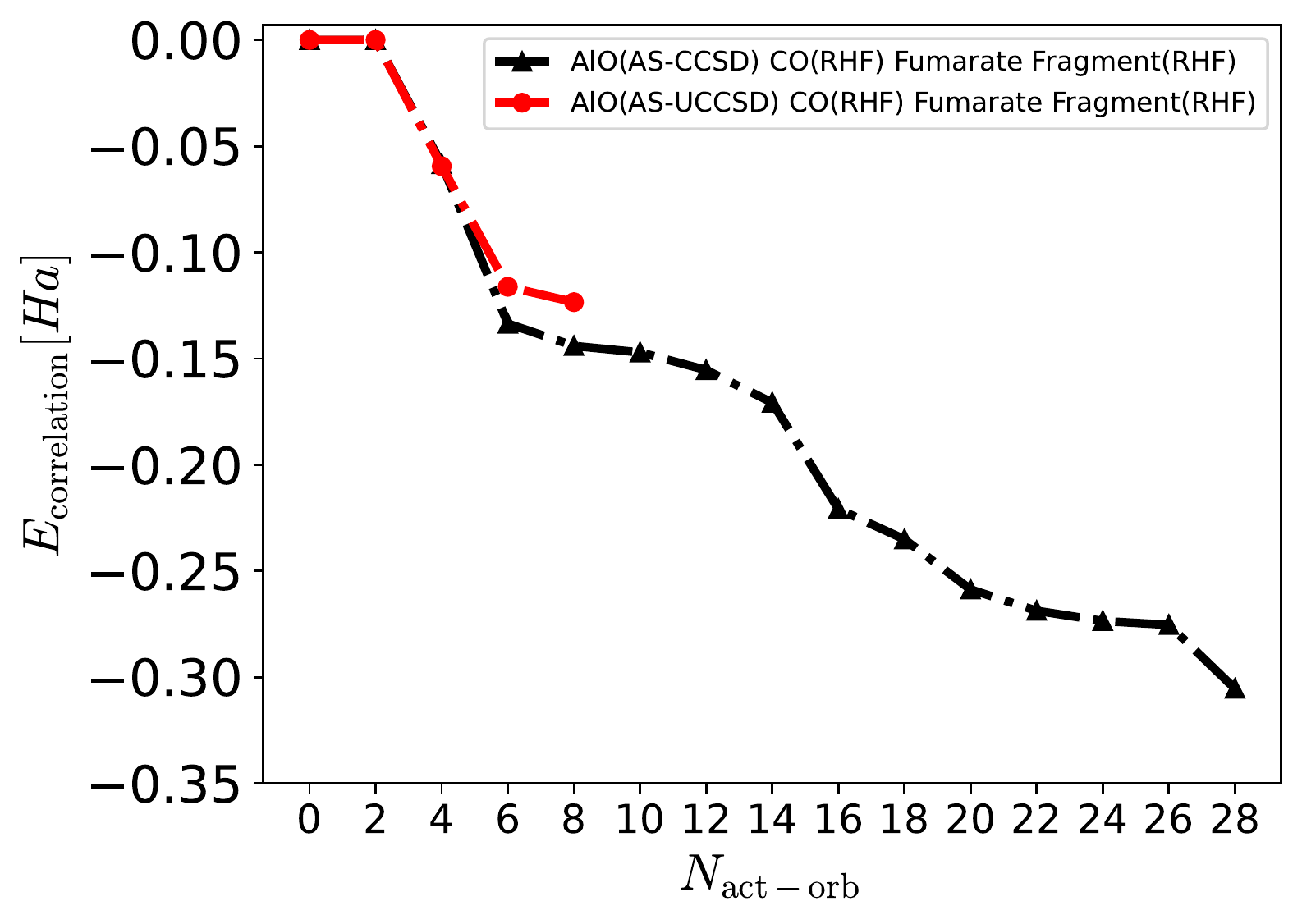}
\caption{Correlation energy versus active space size for CCSD and UCCSD fragment solvers, using the fragmentation depicted in Fig.\ref{fig:sm_edis_frags}d. $N_{\rm act-orb}$ refers to the number of active molecular spatial orbitals.}
\label{dmet_ecorr_vs_norbs_rest_alo_co}
\end{figure} 
\begin{figure}[!h]
\centering
\includegraphics[width=9cm]{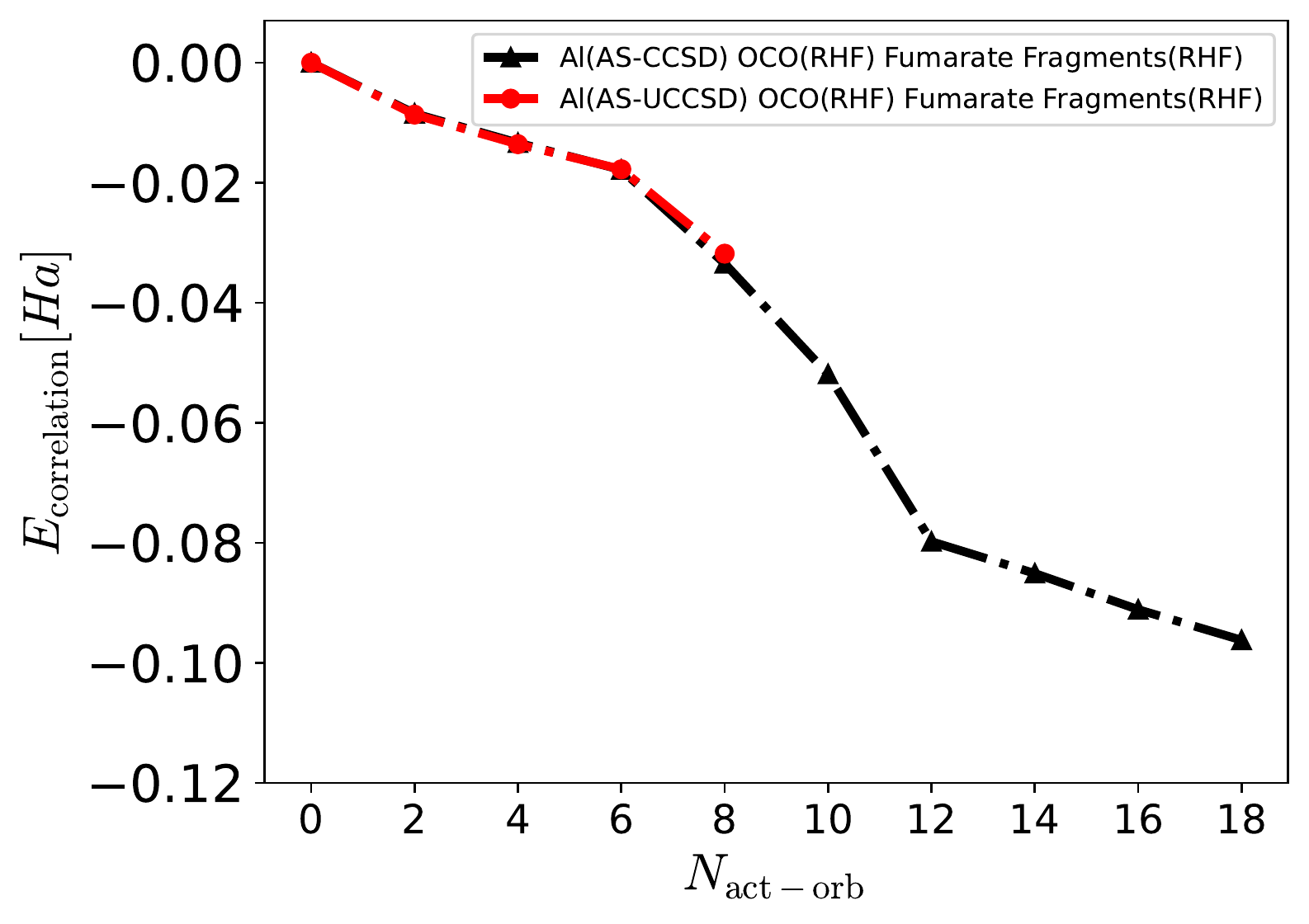}
\caption{Correlation energy versus active space size for CCSD and UCCSD fragment solvers, using the fragmentation depicted in Fig.\ref{fig:Edis}a. $N_{\rm act-orb}$ refers to the number of active molecular spatial orbitals.}
\label{dmet_ecorr_vs_norbs_restsplit_al_oco}
\end{figure} 
\begin{figure}[!h]
\centering
\includegraphics[width=9cm]{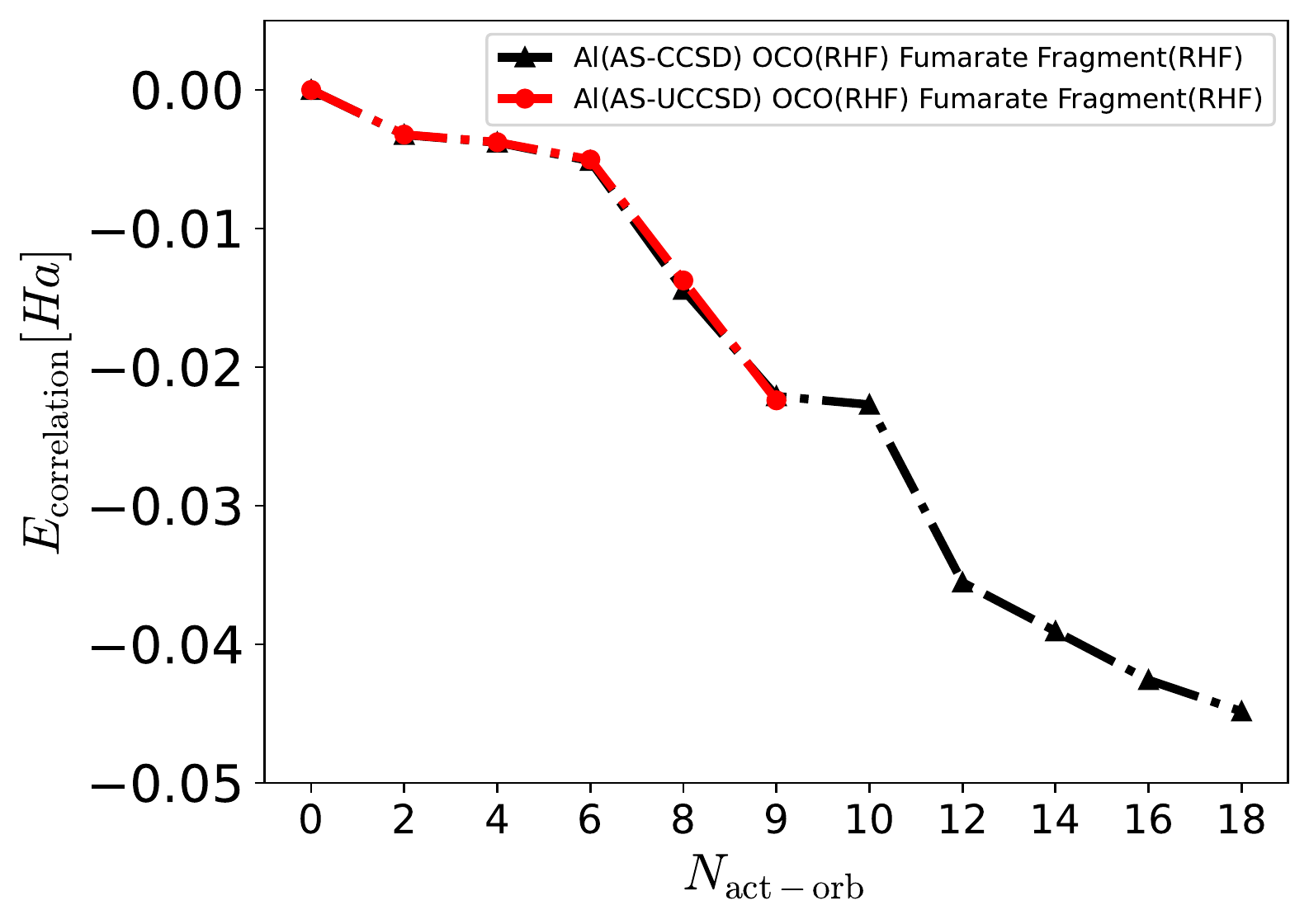}
\caption{Correlation energy versus active space size for CCSD and UCCSD fragment solvers, using the fragmentation depicted in Fig.\ref{fig:Edis}b. $N_{\rm act-orb}$ refers to the number of active molecular spatial orbitals. For $N_{\rm act-orb}$ = 9, 4 occupied and 5 virtual orbitals are used.}
\label{dmet_ecorr_vs_norbs_rest_al_oco}
\end{figure}

In this section, the correlation contribution to the total DMET energy is plotted as a function of active space (AS) size, for all the fragmentations strategies discussed in this work. CCSD and UCCSD are shown for comparison in Figs. \ref{dmet_ecorr_vs_norbs_restsplit_alo_co} - \ref{dmet_ecorr_vs_norbs_rest_al_oco}. Note that the contribution of correlation for full active spaces is much larger in Figs. \ref{dmet_ecorr_vs_norbs_restsplit_alo_co} and \ref{dmet_ecorr_vs_norbs_rest_alo_co} compared to Figs. \ref{dmet_ecorr_vs_norbs_restsplit_al_oco} and \ref{dmet_ecorr_vs_norbs_rest_al_oco}, due to the larger number of orbitals contributed by the O$_{CO_2}$ atom in the AlO fragment.

In Fig. \ref{dmet_ecorr_vs_norbs_rest_alo_co} it can be seen that for 6 active spatial orbitals, the UCCSD correlation energy is approximately 20 mHa smaller than CCSD for a comparable AS. This is consistent with the difference in total energies between UCCSD (N$_q$ = 12) and CCSD with 3 HOMO and 3 LUMO spatial orbitals at Al-O$_{CO_2}$ distance $r = 2\,$\AA. The agreement between UCCSD and CCSD correlation energies does not improve for a larger AS using the fragmentations from Fig. \ref{fig:sm_edis_frags}a,d. At variance, the agreement in correlation energy between UCCSD and CCSD is much better in Figs. \ref{dmet_ecorr_vs_norbs_restsplit_al_oco} and \ref{dmet_ecorr_vs_norbs_rest_al_oco}. This translates to better agreement in dissociation curves for a given AS size (the AS-CCSD dissociation curves are not shown in the main text for brevity). Hence the fragmentations from Fig. \ref{fig:Edis}a and Fig. \ref{fig:Edis}b exhibit better agreement between UCCSD and CCSD for feasible active space sizes in addition to physical dissociation curves.

\section{Classical calculations of Al-Fumarate+CO$_2$}\label{appendix:clas_calc}

\begin{figure}[!h]
\centering
\includegraphics[width=0.55\textwidth]{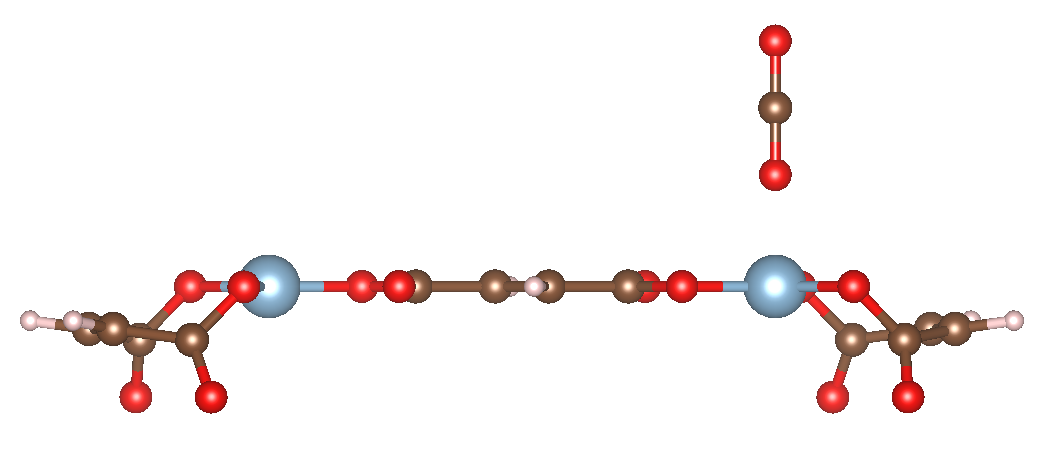}
\caption{Perpendicular orientation of CO$_2$ molecule relative to the Al-fumarate molecule, corresponding to CO$_2$ angle = 90\degree.}
\label{molecules_perp}
\end{figure}
\begin{figure}[!h]
\centering
\includegraphics[width=0.55\textwidth]{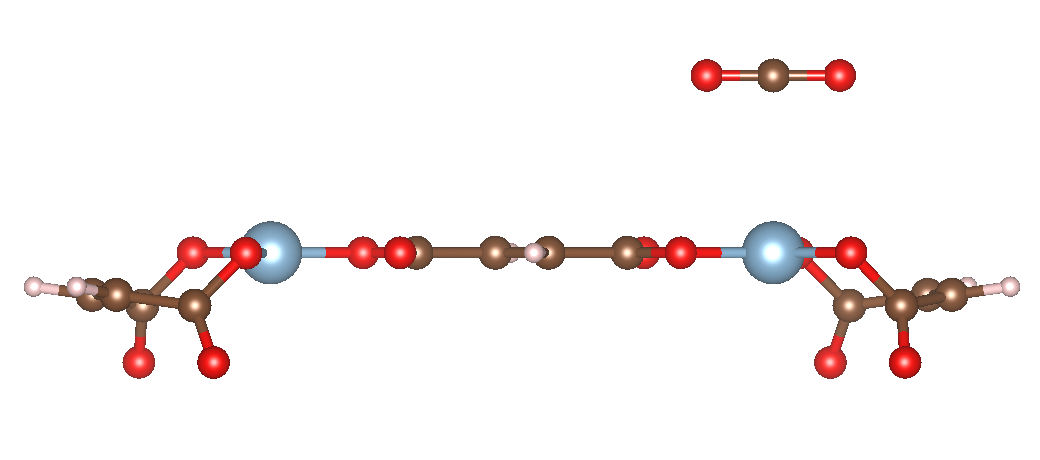}
\caption{Flat orientation of CO$_2$ molecule relative to the Al-fumarate, CO$_2$ angle = 0\degree.}
\label{molecules_flat}
\end{figure}
\begin{figure}[!h]
\centering
	\begin{minipage}{0.55\textwidth}
	\begin{overpic}[width=1.0\textwidth]{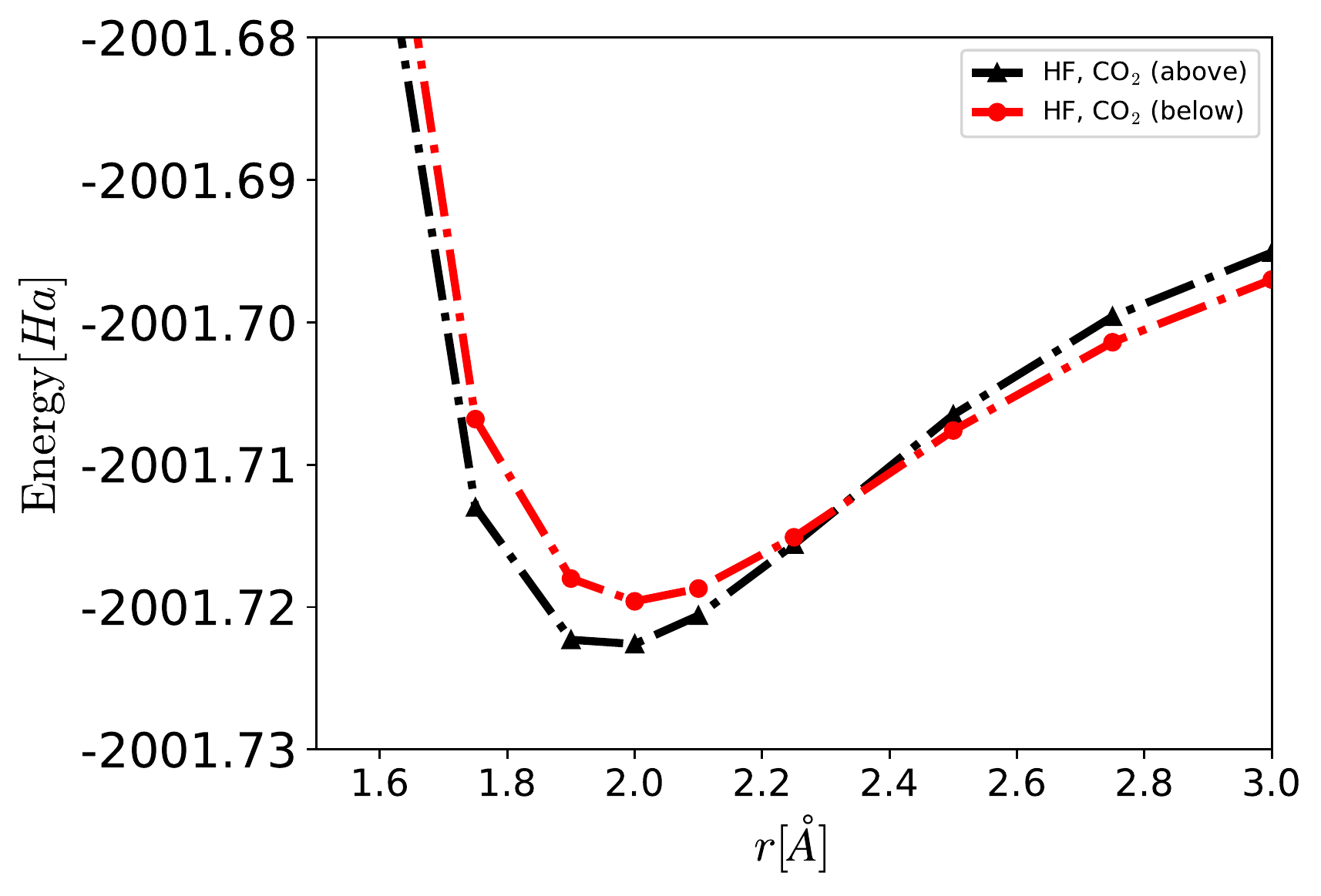}
	\end{overpic}
	\begin{overpic}[width=1.0\textwidth]{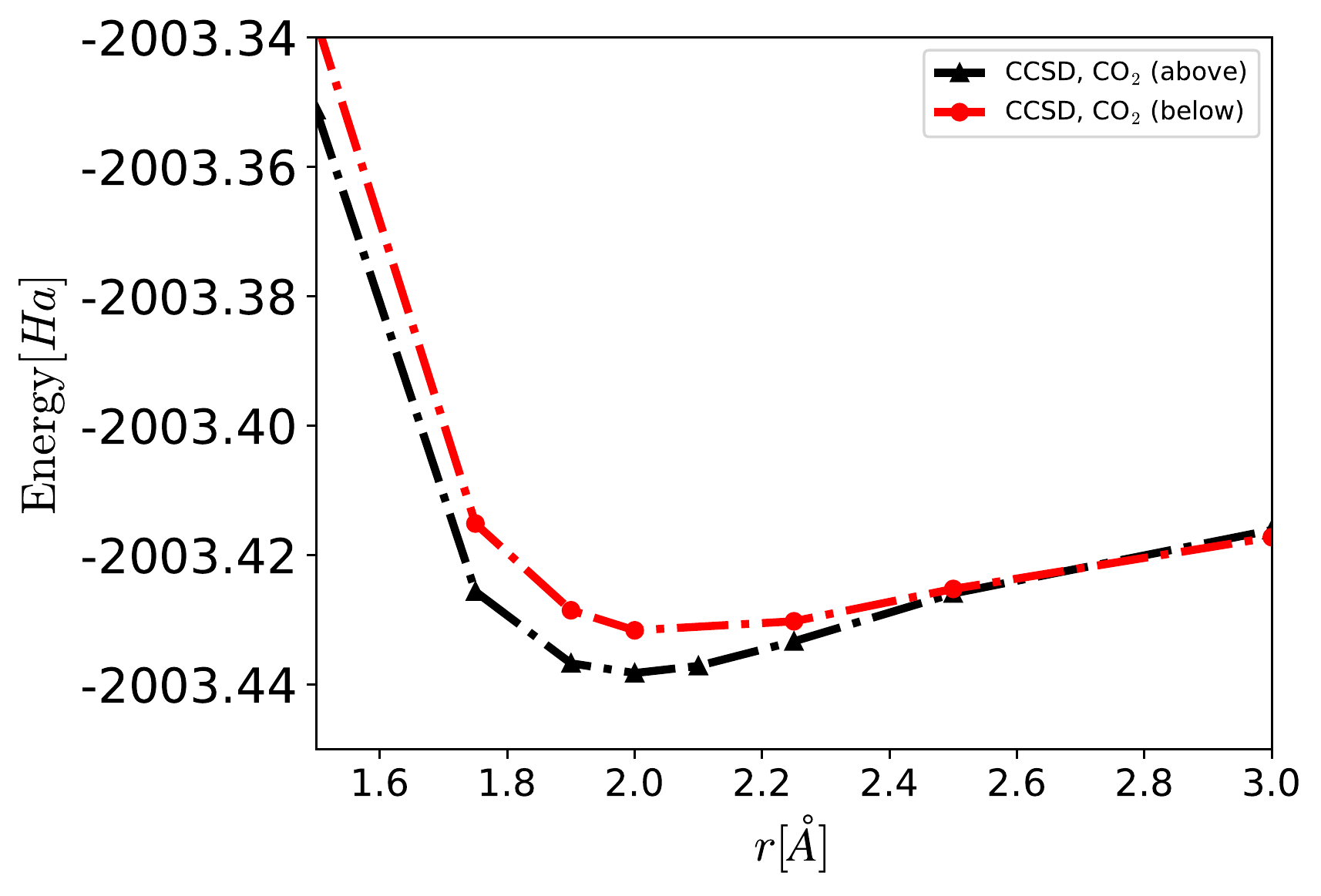}
	\end{overpic}
	\end{minipage}\hspace{1cm}
	
\caption{Total energy of the Al-fumarate+CO$_2$ complex as a function of Al-O distance, for the CO$_2$ molecule incident from ``above'' (as in Fig. \ref{molecules_perp}), or from the opposite direction ``below''. The latter corresponds to increased interaction between CO$_2$ and the fumarate oxygens. No dispersion correction has been added.}
\label{e_versus_dist_abovebelow_nodisp}
\end{figure}

\begin{figure}[!h]
\centering
	\begin{minipage}{0.55\textwidth}
	\begin{overpic}[width=1.0\textwidth]{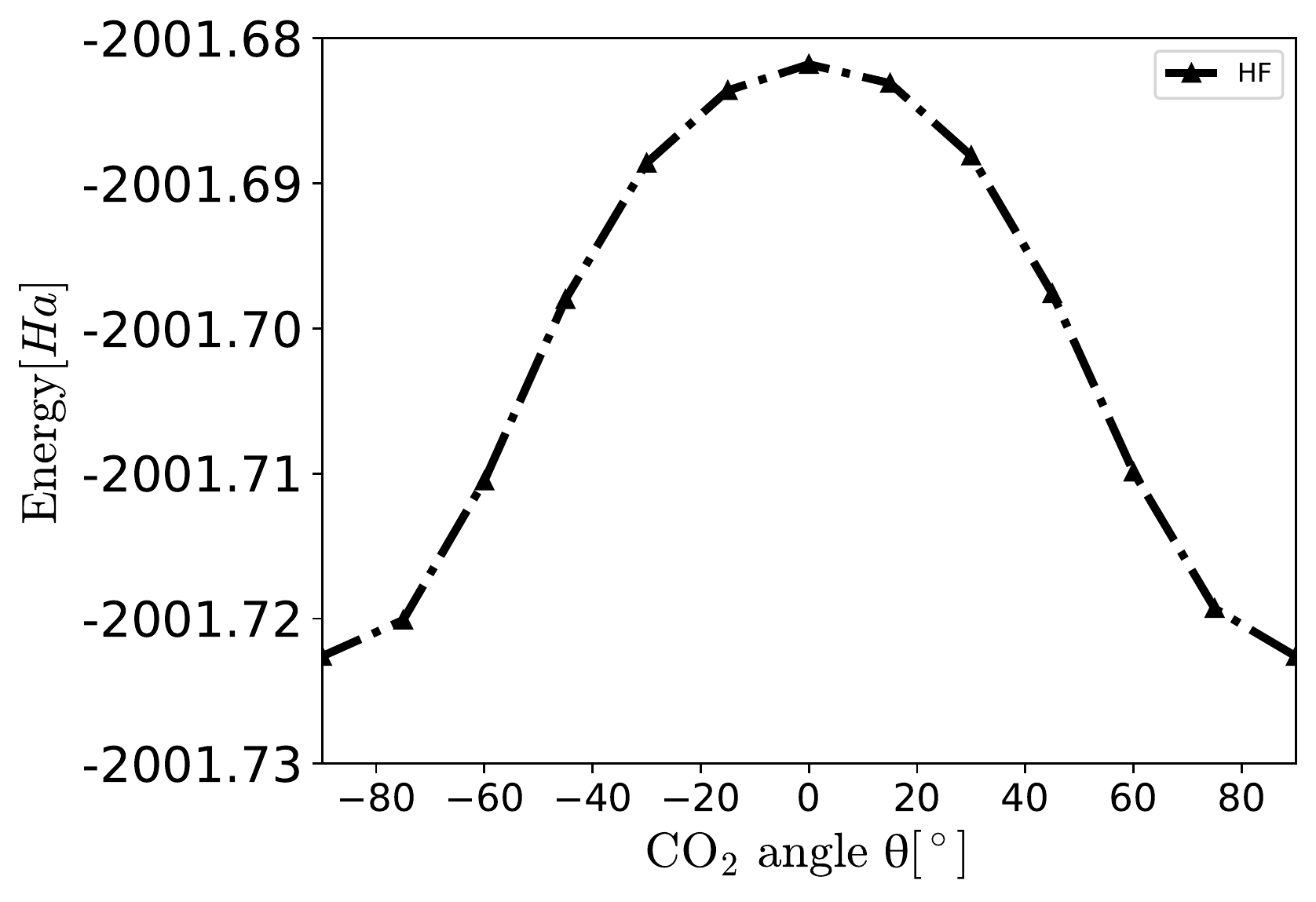}
	\end{overpic}
	\begin{overpic}[width=1.0\textwidth]{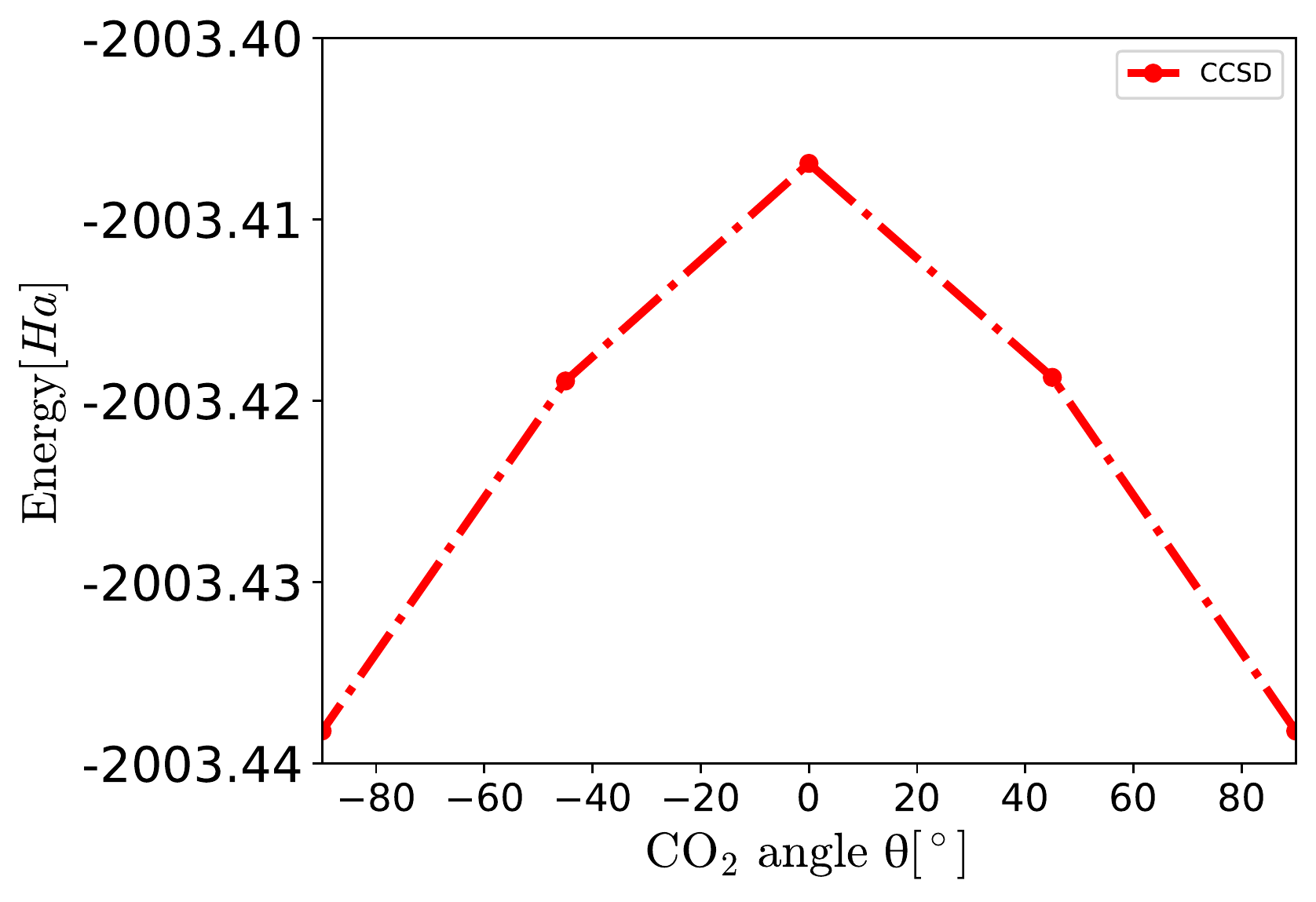}
	\end{overpic}
	\end{minipage}\hspace{1cm}
\caption{Total energy of the Al-fumarate+CO$_2$ complex as a function of CO$_2$ orientation. Distance between Al and C (of CO$_2$) corresponds to 3.197\AA.}
\label{e_versus_angle}
\end{figure}

\begin{figure}[!h]
\centering
\includegraphics[width=0.8\textwidth]{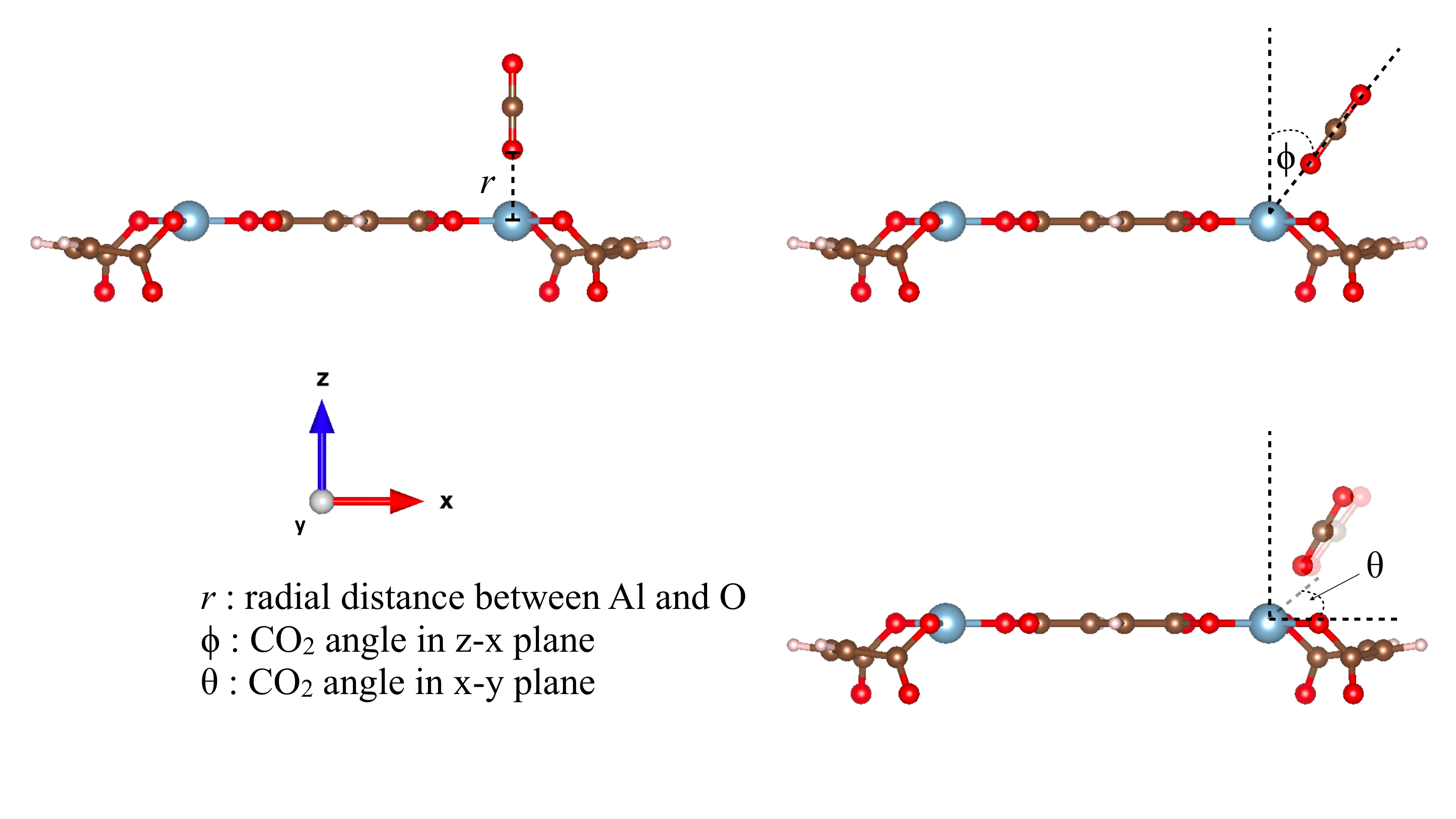}
\caption{Solid angles of CO$_2$ incident on the Al center of the fumarate.}
\label{solid_angle_fumco2}
\end{figure}
\begin{figure}[!h]
\centering
\includegraphics[width=0.7\textwidth]{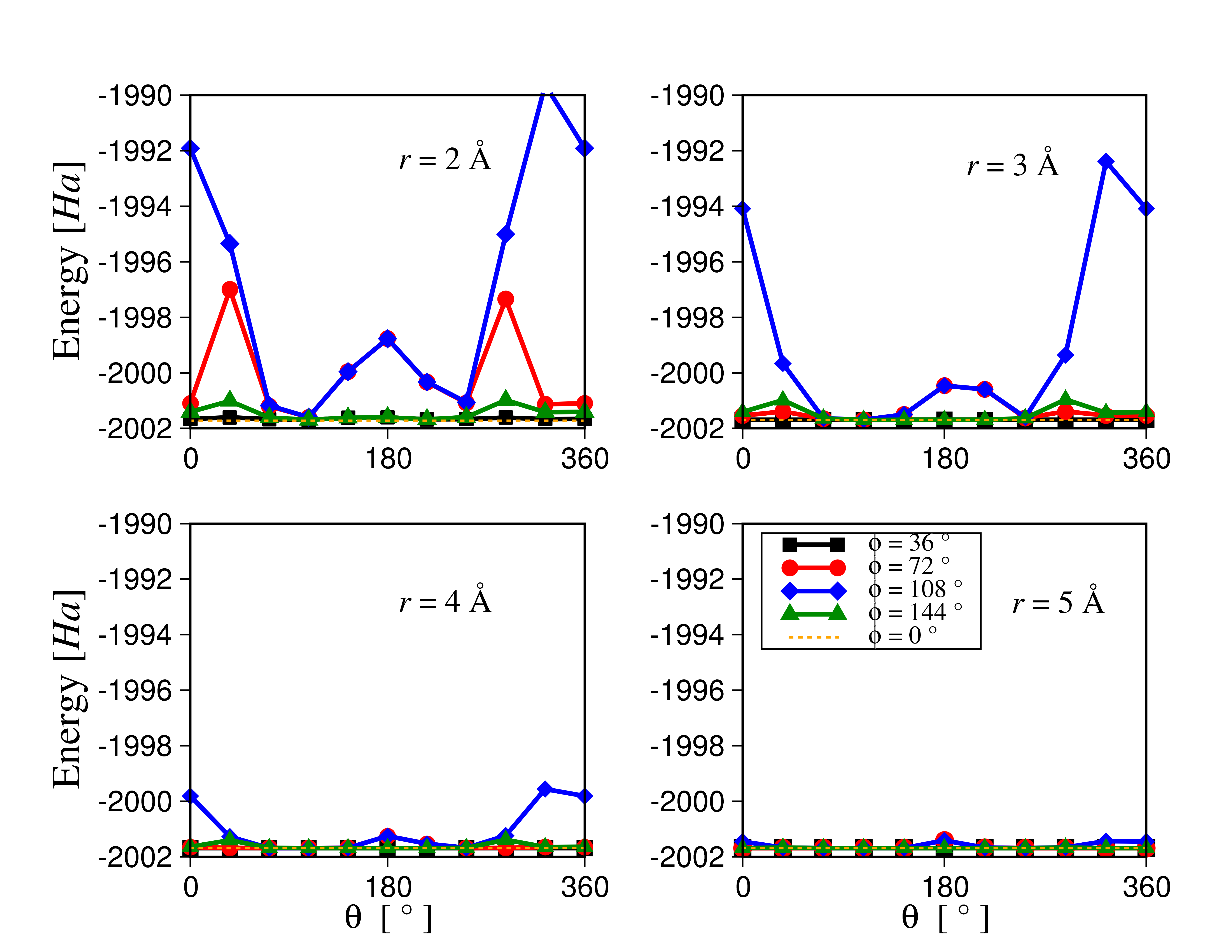}
\caption{Total energy of the Al-fumarate+CO$_2$ complex as a function of CO$_2$ solid incidence angle. For all geometries considered, the CO$_2$ molecule at perpendicular incidence from ``above'' the fumarate plane, with the fumarate Al facing the CO$_2$ O atom, corresponds to the minimum energy.}
\label{e_versus_solidangle}
\end{figure}

 In order to determine the optimal position of the CO$_2$ molecule interacting with the fumarate molecule, a series of classical calculations at various levels of theory (including mean-field and wavefunction techniques) were carried out in which a CO$_2$ molecule was placed at varying distances and orientations relative to the fumarate. In these calculations, the internal geometries of the constituents are kept frozen to their isolated configurations. The lack of inversion symmetry with respect to the horizontal molecular plane prompts the investigation of the lowest energy incidence angle of CO$_2$ as it approaches the Al center. To this end, we calculate the energy as a function of Al-O$_{\mathrm{CO_2}}$ distance at perpendicular orientation (relative to the horizontal plane), for the CO$_2$ molecule approaching the fumarate from ``above'' and ``below'' the horizontal plane (where ``above'' corresponds to Figs. \ref{molecules_perp} and \ref{molecules_flat}). Results are plotted in Fig. \ref{e_versus_dist_abovebelow_nodisp}. Following this, Fig. \ref{e_versus_angle} shows the energy as a function of CO$_2$ angle relative to the fumarate plane (keeping the Al-C$_{\mathrm{CO_2}}$ distance fixed), in which the CO$_2$ is placed ``above'' the fumarate plane.

The dependence of fumarate+CO$_2$ energy on CO$_2$ incidence angle was also evaluated for the purpose of determining the optimal geometry. Solid angle parameters are defined as in Fig. \ref{solid_angle_fumco2}, while in Fig. \ref{e_versus_solidangle} the energy as a function of solid angle coordinates for the CO$_2$ molecule is reported at the mean-field Hartree-Fock level. This further indicates that the minimum energy geometry for the minimal STO-3G basis set corresponds to the CO$_2$ at perpendicular incidence from ``above'' the fumarate plane (see Fig. \ref{molecules_perp}) with the Al-O$_{\mathrm{CO_2}}$ distance at approximately 2 \AA.\\

\section{Quantum Computational DMET in detail}\label{sec:DMET}

If the exact ground state wavefunction of the full system is known one can construct a projector $\hat{P}^x$ for fragment $x$ such that the projected reduced size Hamiltonian (embedding Hamiltonian) $\hat{H}^x = \hat{P}^x \hat{H} \hat{P}^x$ results in the same exact ground state\cite{knizia13, wouters16}. As the exact wavefunction is in general unknown, the embedding Hamiltonian is constructed with a wavefunction approximated by a low level theory such as HF.

The DMET algorithm starts by constructing the Hamiltonian ($\hat{H}$) with a localised and orthogonal basis, in which the domain of each fragment can be specified. Then the algorithm solves the full problem with the HF theory and calculates the one-electron reduced density matrix (1-RDM) in the localised basis. More precisely, a modified total Hamiltonian $\hat{H}'$ is solved with HF theory,   

\begin{equation} \label{ham_plus_u}
\hat{H'} = \hat{H} + \sum_{x} \sum_{ij \in A^{x}} u^{x}_{ij} a^\dagger_i a_j 
\end{equation}

 where $A^{x}$ refers to the subspace spanned by fragment $x$, and $a^{\dagger}_i$ is a fermionic creation operator in the localised basis. The extra one-body term in Eq. \ref{ham_plus_u} is the correlation potential that accounts for the effects of correlations on the 1-RDM. The values of $u^{x}_{ij}$ are to be determined such that the 1-RDM on each fragment block matches the 1-RDM calculated with the high level methods, and they are improved self-consistently as described by Wouters et al.\cite{wouters16}. In this paper we perform only the one-shot DMET method, which is equivalent with the first iteration of DMET when the initial correlation potential is zero, thus $\hat{H}' = \hat{H}$.

Following the calculation of the one-electron reduced density matrix of the full system, a fragment projector $\hat{P}^x$ can be constructed based on the HF solution and the embedding Hamiltonian $\hat{H}^x$ can be expressed \cite{wouters16}. In practice, the Schmidt decomposition of the HF wavefunction allows one to partition the 1-RDM into a fragment and its complementary subsystems, and by diagonalising the complementary sub-block of the 1-RDM one can construct a useful basis (Schmidt basis) in which the fractionally occupied orbitals are kept as bath orbitals and the occupied and empty orbitals are designated as the environment. The embedding Hamiltonian in the Schmidt basis is

\begin{equation}
    \hat{H}^x(\mu_{\rm global}) =  \sum_{ij \in A^x \cup B^x} \Bigg(h^x_{ij} + \sum_{kl \in N} (V^x_{ijkl} - V^x_{ilkj}) \boldsymbol{D}_{kl}^{\text{env},x}\Bigg) \hat{c}^\dagger_{i} \hat{c}_j +
    \frac{1}{2}\sum_{ijkl \in A^x \cup B^x} V^x_{ijkl} \hat{c}^\dagger_i \hat{c}_k^\dagger \hat{c}_{l} \hat{c}_j - \mu_{\rm global} \hat{N}_x  
\label{hamx}
\end{equation}

 where $h^x_{ij}$ and $V^x_{ijkl}$ are the one and two-electron integrals in the Schmidt basis, respectively, $c^{\dagger}_i$ is the corresponding fermionic creation operator, and $B^x$ refers to the subspace of the bath orbitals. The $\boldsymbol{D}_{kl}^{\text{env},x}$ is the one-electron reduced density matrix of the fully occupied orbitals in the environment. The last term is not the result of the projection, but it is added to control the charge distribution between the fragment and the bath. $\hat{N}_x=\sum_{i \in A^x} \hat{c}^\dagger_{i} \hat{c}_i$ is the particle number operator for fragment $x$ and $\mu_{\rm global}$ is the global chemical potential, independent of the fragment, and is determined from the constraint

\begin{equation}
    \sum_{x} \langle\Psi_{x}(\mu_{\rm global}) | \hat{N}_{x} | \Psi_{x}(\mu_{\rm global})\rangle = N_{e}
\end{equation}

where $\Psi_{x}(\mu_{\rm global})$ is the ground state of the $\hat{H}^x(\mu_{\rm global})$ obtained with a high level method, such as VQE, and $N_e$ is the total number of electrons in the molecule. In practice, this constraint is satisfied iteratively, with initial value $\mu_{\rm global} = 0$. Once the final value of $\mu_{\rm global}$ and $\Psi_{x}(\mu_{\rm global})$ are found the energy of each fragment is obtained (assuming ``democratic'' mixing of local RDMs \cite{wouters16}) from 

\begin{equation} \label{e_frag}
E^x = \sum_{i \in A^x} \Bigg[ \sum_{j \in A^x \cup B^x} \Bigg(h^x_{ij} + \frac{1}{2} \sum_{kl \in N} (V^x_{ijkl} - V^x_{ilkj}) \boldsymbol{D}_{kl}^{\text{env},x}\Bigg)\boldsymbol{P}^x_{ij} + \frac{1}{2}\bigg(\sum_{jkl \in A^x \cup B^x}V^x_{ijkl}\boldsymbol{\Gamma}^x_{ijkl}\bigg) \Bigg]
\end{equation}

where $\boldsymbol{P}^x_{ij}=\langle\Psi_{x}(\mu_{\rm global}) | \hat{c}^\dagger_{i} \hat{c}_j | \Psi_{x}(\mu_{\rm global})\rangle$ and 
$\boldsymbol{\Gamma}^x_{ijkl}=\langle\Psi_{x}(\mu_{\rm global}) | \hat{c}^\dagger_i \hat{c}_k^\dagger \hat{c}_{l} \hat{c}_j  | \Psi_{x}(\mu_{\rm global})\rangle$.

In Quantinuum's computational chemistry platform, one may choose to calculate the density matrix for each fragment either classically or with one of the available quantum algorithms. This enables a mixing of algorithms for different fragments: less important regions of the molecule can be solved with a cheap classical method, while the important fragments where the interesting chemistry occurs (in this case the Al-CO$_2$ interaction) may be treated with high accuracy wavefunction methods. The latter can include a quantum computational solver for the fragment of interest, which allows for the combination of classical and quantum calculation methods for different parts of the molecule. In addition, we can choose active orbital spaces for different fragments, which allows for more efficient simulations in terms of the number of qubits and number of interaction terms. Once all these quantities have been calculated, the total energy of the molecule will consist of a sum of fragment energies $E^{x}$ plus the nuclear repulsion contribution

\begin{equation} \label{dmet_etot}
E = \sum_{x} E^{x} + E_{nuc} .
\end{equation}

While many embedding approaches are available \cite{sun16}, DMET is an approach that has been shown to accurately describe the dissociation of strongly correlated systems whose dissociated state is in general difficult to capture, even when the system is fragmented to individual atoms \cite{knizia13}. In this work, the energetics of CO$_2$ dissociation from the fumarate will be investigated as a means to study the Al-CO$_2$ interaction, and for this reason an embedding approach which is expected to capture the strong coupling between the embedded fragment and its environment, as well as the static correlation resulting from the dissociation of the fragmented system, is desired. Hence, we select the DMET approach for this purpose. However, despite these previous works \cite{knizia13}, our results show that the predicted dissociation behavior is highly dependent on the adopted fragmentation strategy. While quantitative differences between different DMET fragmentations of the same system have been reported recently \cite{wouters16, li21}, our work shows that different fragmentations can result in \textit{qualitative} differences in the predicted behavior of the total system, even leading to differing predictions as to whether a chemical complex is bound or unbound.

The variational quantum eigensolver (VQE) \cite{peruzzo} is a hybrid quantum/classical algorithm that can variationally solve for the ground state (or other eigenstates, given appropriate constraints) of a given Hamiltonian. We use VQE as the parameter optimiser of the UCCSD wavefunction ansatz. The latter has been implemented to support active orbital spaces, which is useful for controlling the number of qubits in the simulation. This comprises our quantum computational fragment solver. Note that the UCCSD wavefunction can be written as

\begin{equation} \label{uccsd_psi}
|\Psi_{\text{UCCSD}}\rangle = \mathcal{U}|\psi_0\rangle
\end{equation}

where $\psi_0$ labels the reference state, typically the HF wavefunction, and $\mathcal{U}$ is the unitary operator constructed from anti-hermitian excitation generators

\begin{equation} \label{unitary}
\mathcal{U} = e^{\sum_i{\theta_i(\hat{T}_i - \hat{T}^{\dagger}_i)}}
\end{equation}

where the fermionic excitations $\hat{T}_i$ are restricted to single and double excitations, and $\theta_i$ are variational parameters. Due to the difficulty of implementing $\mathcal{U}$ directly in the NISQ era, it is typically decomposed into an ordered product of tractable operations, which is known as Trotter decomposition \cite{suzuki76}

\begin{equation} \label{trotter_unitary}
e^{\sum_i{\theta_i(\hat{T}_i - \hat{T}^{\dagger}_i)}} \approx \Big[\prod_i e^{\frac{\theta_i(\hat{T}_i - \hat{T}^{\dagger}_i)}{t}}\Big]^t .
\end{equation}

This is an approximate relation since the generators in the exponent do not commute in general. The first-order Trotter approximation, utilised in Quantinuum's computational chemistry platform, is obtained when $t = 1$. In our implementation, we order the terms such that single excitations are applied first, followed by double excitations. Within the singles and doubles, orbital indexes are ordered such that lowest occupied and virtual orbitals are applied first. We use the same basis sets as the classical calculations to generate the integrals for the UCCSD calculations.

In order to run 16 or more qubit simulations using a classically simulated quantum hardware, high performance computational resources are required. We utilise Microsoft's Azure for idealised noise-free simulations. This allows for the benchmarking of this method without the interference of hardware noise. Hence, results from idealised calculations are used to study the intrinsic accuracy of this approach. The IBMQ emulator for \textit{ibm\_lagos} is used for simulations with a noise model (the calibration data for this noise model is available from the authors on request). This allows for the effect of noise on calculations of a dissociation barrier, and the impact of the choice of error mitigation scheme. Thus our work sheds light on the theoretical accuracy of this approach,  as well as the effect of hardware error and its mitigation, for systems relevant to carbon capturing MOFs which will serve as a guide to future simulations on fault-tolerant quantum devices. 
\end{document}